\documentclass[sigconf]{acmart}

\AtBeginDocument{%
  \providecommand\BibTeX{{%
    \normalfont B\kern-0.5em{\scshape i\kern-0.25em b}\kern-0.8em\TeX}}}

\copyrightyear{2024}
\acmYear{2024}
\setcopyright{acmlicensed}\acmConference[UIST '24]{The 37th Annual ACM Symposium on User Interface Software and Technology}{October 13--16, 2024}{Pittsburgh, PA, USA}
\acmBooktitle{The 37th Annual ACM Symposium on User Interface Software and Technology (UIST '24), October 13--16, 2024, Pittsburgh, PA, USA}
\acmDOI{10.1145/3654777.3676386}
\acmISBN{979-8-4007-0628-8/24/10}

\usepackage{acmart-taps}
\usepackage{amsmath}
\usepackage{multirow}  
\usepackage{tabularx}
\usepackage{makecell}
\usepackage{longtable}
\usepackage{colortbl}
\usepackage[ruled,linesnumbered]{algorithm2e}

\usepackage{multicol}  
\usepackage{array}
\usepackage{xtab}
\makeatletter
\patchcmd{\estimate@lineht}{1\p@}{-1.5\p@}{}{}
\makeatother

\begin{document}

\title{VisionTasker: Mobile Task Automation Using Vision Based UI Understanding and LLM Task Planning}

\author{Yunpeng Song}
\orcid{0000-0002-4186-0408}
\affiliation{%
  \institution{MOE KLINNS Lab}
  \country{Xi'an Jiaotong University}
}
\email{yunpengs@xjtu.edu.cn}

\author{Yiheng Bian}
\orcid{0009-0004-8628-1751}
\affiliation{%
  \institution{MOE KLINNS Lab}
  \country{Xi'an Jiaotong University}
}
\email{yhbian@stu.xjtu.edu.cn}

\author{Yongtao Tang}
\orcid{0009-0007-9941-8456}
\affiliation{%
  \institution{MOE KLINNS Lab}
  \country{Xi'an Jiaotong University}
}
\email{tyt18050701377@stu.xjtu.edu.cn}

\author{Guiyu Ma}
\orcid{0009-0000-9910-5301}
\affiliation{%
  \institution{MOE KLINNS Lab}
  \country{Xi'an Jiaotong University}
}
\email{guiyu.ma@stu.xjtu.edu.cn}

\author{Zhongmin Cai}
\authornote{Corresponding author.}
\orcid{0000-0002-4903-3992}
\affiliation{%
  \institution{MOE KLINNS Lab}
  \country{Xi'an Jiaotong University}
}
\email{zmcai@sei.xjtu.edu.cn}

\renewcommand{\shortauthors}{Song et al.}

\begin{abstract}
Mobile task automation is an emerging field that leverages AI to streamline and optimize the execution of routine tasks on mobile devices, thereby enhancing efficiency and productivity. Traditional methods, such as Programming By Demonstration (PBD), are limited due to their dependence on predefined tasks and susceptibility to app updates. Recent advancements have utilized the view hierarchy to collect UI information and employed Large Language Models (LLM) to enhance task automation. However, view hierarchies have accessibility issues and face potential problems like missing object descriptions or misaligned structures. This paper introduces VisionTasker, a two-stage framework combining vision-based UI understanding and LLM task planning, for mobile task automation in a step-by-step manner. VisionTasker firstly converts a UI screenshot into natural language interpretations using a vision-based UI understanding approach, eliminating the need for view hierarchies. Secondly, it adopts a step-by-step task planning method, presenting one interface at a time to the LLM. The LLM then identifies relevant elements within the interface and determines the next action, enhancing accuracy and practicality. Extensive experiments show that VisionTasker outperforms previous methods, providing effective UI representations across four datasets. Additionally, in automating 147 real-world tasks on an Android smartphone, \textcolor{black}{VisionTasker demonstrates advantages over humans in tasks where humans show unfamiliarity} and shows significant improvements when integrated with the PBD mechanism. \textcolor{black}{VisionTasker is open-source and available at \url{https://github.com/AkimotoAyako/VisionTasker}.}
\end{abstract}

\ccsdesc[500]{Human-centered computing~Graphical user interfaces}

\keywords{Mobile Task Automation, Large Language Models}

\maketitle

\section{Introduction}

Smartphones have become indispensable in our daily routines, handling tasks like managing emails, paying bills, and ordering food. However, for individuals with sensory or motor limitations, such as restricted hand mobility, or in scenarios demanding divided attention like driving, smartphones can become less accessible despite their intuitive interfaces. Challenges also arise in repetitive tasks, such as setting up numerous calendar entries or sending similar messages to multiple contacts, which are quite common in both daily usage and mobile app testing scenarios. Additionally, interacting with different Apps can be confusing and prone to errors for elderly users, particularly when navigating complex menu systems. Consequently, investigating methods to augment smartphone intelligence for task automation that aligns with user intent is a valuable pursuit, promising significant enhancements in usability and productivity.

Previous research has explored mobile task automation through Programming By Demonstration(PBD) \cite{li2017sugilite,li2018kite,sereshkeh2020vasta,riva2021etna,vu2023voicify}. This technique records user actions to create scripts for later use. However, PBD is restricted to tasks users have previously demonstrated and necessitates manually defining task intents and parameters for modification. Updates to apps, which alter workflows or interfaces, may largely affect the effectiveness of PBD. Researchers have also delved into training a multi-modal model to decipher the relationship between user commands and actions based on large-scale datasets in an end-to-end manner \cite{li2020mapping,sun2022meta,li2023spotlight,zhan2023you}. Yet, constructing large, high-quality datasets is challenging, and current datasets frequently contain human trials in task execution, which are random and largely irrelevant to the executed task, leading to many incorrect samples. Furthermore, the task planning skills that models acquire from these datasets are usually confined to known instructions and interfaces, showing a lack of adaptability to new situations. More recent approaches have leveraged the reasoning abilities of Large Language Models to enhance task automation~\cite{wang2023enabling,wen2023empowering}. Prior works primarily employ view hierarchies to gather UI information. However, view hierarchies are not always accessible and are prone to issues like missing object descriptions or misaligned structures.

Considering the variety of tasks and the flexibility of app interfaces and workflows, let's step back and examine how humans typically complete tasks using mobile apps. Effective app design hinges on intuitive interaction, encapsulating principles such as "recognition rather than recall"~\cite{nielsen1994enhancing}. This principle suggests that users should be able to easily understand and use an app, even a new one, based primarily on UI cues.
Additionally, humans often tackle complex problems by dividing them into smaller sequential subtasks. Take, for example, the process of paying an electricity bill using a mobile app. A user, perhaps unfamiliar with the app, would start by exploring its UI. She then may find a section labeled ``City Services'', which seems relevant to bill payment. Delving into it, they may next navigate to ``Utilities'' among various presented services, and finally locate and select the option for electricity bill payments. This process reflects step-by-step reasoning, where the user focuses on identifying relevant elements from the current UI and utilizing them to advance the task. Inspired by this approach, we pose the question: Is it possible to automate mobile tasks through a human-like strategy of sequential visual recognition and planning when interacting with interfaces?

To address this, we introduce VisionTasker, a two-stage framework designed for task automation that emphasizes vision-based UI understanding and task planning.
The first stage involves analyzing a UI screenshot and extracting semantic information. This includes building models to detect UI elements, recognize text, group them into semantically meaningful blocks based on visual layout, and output the UI interpretation using the natural language. In the second stage, we employ LLMs for task planning. Since directly generating action sequences to complete the task can lead to impractical instructions due to ``hallucination''~\cite{maynez-etal-2020-faithfulness}. Utilizing the capability of vision based UI understanding, we introduce a step-by-step approach, presenting a description of current interface image at each step of the task execution to the LLM and asking it to identify relevant elements within the interface and determine the next move. This strategy not only provides context but also ensures decisions are based on actual UI elements, leading to more accurate instructions. Additionally, we incorporate a Programming By Demonstration (PBD) mechanism to assist the LLM in complex task planning situations.

We implement our approach using public GUI datasets for training the UI understanding models, and the off-the-shelf ERNIE Bot as LLM navigator. We conduct extensive experiments to evaluate the performance and challenges of our method, and the results demonstrate that, VisionTasker, our vision-based UI understanding scheme surpasses previous methods, offering effective UI representations across four public datasets. Moreover, our method exhibits impressive performance, comparable to that of human abilities, in automating 147 real-world tasks on an Android smartphone. \textcolor{black}{It shows advantages over humans in handling unfamiliar tasks and significant improvement when combined with the PBD mechanism.} The main contributions of this paper are as follows:
\begin{itemize}
    \item We propose VisionTasker, a two-stage framework combining vision-based UI understanding and LLM task planning, for mobile task automation in a step-by-step manner. Our approach eliminates the need for accessing view hierarchies to represent the UI and a vast dataset to train large models.
    \item We tailor a scheme of vision based UI understanding and devise a systematic procedure to transform the UI screenshot into expressive UI layout semantics in the form of natural language. The scheme can be further enhanced by incorporating the Programming by Demonstration (PBD) mechanism. 
    \item Through comprehensive experiments, we demonstrate the effectiveness and challenges of our approach in automating tasks across screen sizes, task types, and levels of complexity. 
\end{itemize}

\section{Related Work}

\textcolor{black}{Past approaches to automating tasks on mobile devices fall into three main categories: first, watching how users interact with their devices to create scripts that can repeat similar actions; second, using large datasets to train custom models that understand screen content and link instructions to UI elements; and third, using LLMs and multimodal LLMs to plan tasks.}

\subsection{Programming by Demonstration}

Programming by Demonstration (PBD) offers an intuitive way to automate tasks on smartphones. This method typically captures users' actions to create scripts that can be activated later. For example, SUGILITE~\cite{li2017sugilite} performs a pioneering study utilizing the Android accessibility API to grasp the app's UI structure and using a conversational UI for script creation and task execution. Appinite~\cite{li2018appinite} extends this by incorporating Natural Language Understanding (NLU) to better recognize UI elements in response to user commands. Further advancements are seen in systems like Etan~\cite{riva2021etna} and KITE~\cite{li2018kite}, which introduce task-oriented bots assisting in template creation for app functions. AutoVCI~\cite{pan2022automatically} improves user guidance by asking questions to pinpoint target actions and parameters. VASTA~\cite{sereshkeh2020vasta} enhances script robustness by processing user queries in natural language, allowing for flexible task descriptions and triggering appropriate scripts. ParamMacros~\cite{krosnick2022parammacros} enables users to generalize queries and identify parameters through text annotation, thus broadening systems' applicability. However, these methods typically rely on precise, user-defined scripts, making them vulnerable to changes in the GUI and workflow from the demonstration to the execution phase. This reliance means only the exact procedures converted into scripts can be replayed, limiting flexibility and adaptability.


\subsection{Understanding the Interfaces}

Previous research has focused on understanding the entire UI screen by converting it into a vector, which is then utilized for tasks like retrieving screens and classifying apps~\cite{li2021screen2vec,Fu2021UnderstandingMG,ang2022learning}. Studies have also taken a more detailed approach, such as extracting information like object attributes and structural details of the UI~\cite{zhang2021screen,wu2021screen,li-etal-2020-widget,xie2022psychologically}, \textcolor{black}{recognizing non-text icons~\cite{chen2020object,chen2022towards,he2021actionbert}, and presenting the entire UI using textual representations~\cite{wang2021screen2words,leiva2022describing,baechler2024screenai}. }

\textcolor{black}{The creation of large-scale datasets linking UIs with text descriptions/instructions has led researchers to focus on training vision-language models for mobile interface understanding~\cite{li2020mapping,sun2022meta,burns2022dataset,venkatesh2022ugif,rawles2023androidinthewild}. These models primarily aim to connect UI elements with specific text instructions.} Li et al.~\cite{li2020mapping} use a dual-transformer approach to link text and images based on a dataset of 295k synthetic single-step UI-instruction pairs. Similarly, ILuvUI~\cite{jiang2023iluvui} uses a large dataset of 353k text-image samples, along with visual and text encoders and a LLM as the decoder, to generate task planning responses. Li et al.~\cite{li2023spotlight} collected a dataset of 2.5 million screens with view hierarchies to train a vision-language model that can determine if a target region matches a one-step natural language command. Rawles et al.~\cite{rawles2023androidinthewild} and Zhan et al.~\cite{zhan2023you} use the AITW dataset, which includes 30k instructions and corresponding UI sequences, to train multimodal transformers that predict the next action needed to complete a task. Yet, these models are heavily shaped by training data, and current datasets frequently contain human trials in task execution, which are largely irrelevant to the target task, leading to many incorrect samples. Besides, the planning capability that models acquire from these datasets are confined to known instructions and UIs, showing a lack of adaptability to adjustments in the UI and processes, e.g., introduced through app updates.

\subsection{Automation with Large Language Models}

Researchers also explore leveraging the generic knowledge and reasoning ability of LLM for automation. This approach typically involves a two-step process: identifying UI elements and using an LLM to infer the elements that require interaction. 
Wang et al.~\cite{wang2023enabling} conduct the pioneering research to transform the view hierarchy into an HTML-like structure for using LLMs to enable conversational interaction on mobile UIs. AutoDroid~\cite{wen2023empowering} combines LLMs and app-specific knowledge in a novel way by generating a UI Transition Graph through offline exploration. 
ResponsibleTA~\cite{zhang2023responsible} leverages LLMs to plan the entire task action sequences, supplementing this with an auxiliary model to ensure the viability, completeness, and security of the plans. Meanwhile, MM-Navigator~\cite{yan2023gpt} employs the SegmentAnything Model (SAM)~\cite{kirillov2023segment} to dissect UI into identifiable segments, each marked with an ID, and combines this with user tasks in GPT-4V to generate relevant instructions. 
\textcolor{black}{Recent work has also applied LLMs to automated GUI testing~\cite{liu2023chatting}, accessibility testing~\cite{taeb2024axnav}, and bug replay~\cite{feng2024prompting}.} These works rely on view hierarchies or general models like SAM to identify UI elements, which can be impractical or yield low data quality. 

\textcolor{black}{Recent advances have also led to multimodal LLMs capable of handling various UI tasks, including answering questions and automating processes. These models, such as Ferret UI~\cite{you2024ferret}, CogAgent~\cite{hong2024cogagent}, and Fuyu~\cite{ADEPT}, take a different approach by combining UI understanding and task planning in a single stage. While these models show promise, their performance still needs improvement, especially in accurately grounding UI elements.}

\section{Overview of Our Method}

\begin{figure}
    \centering
    \includegraphics[width=1\linewidth]{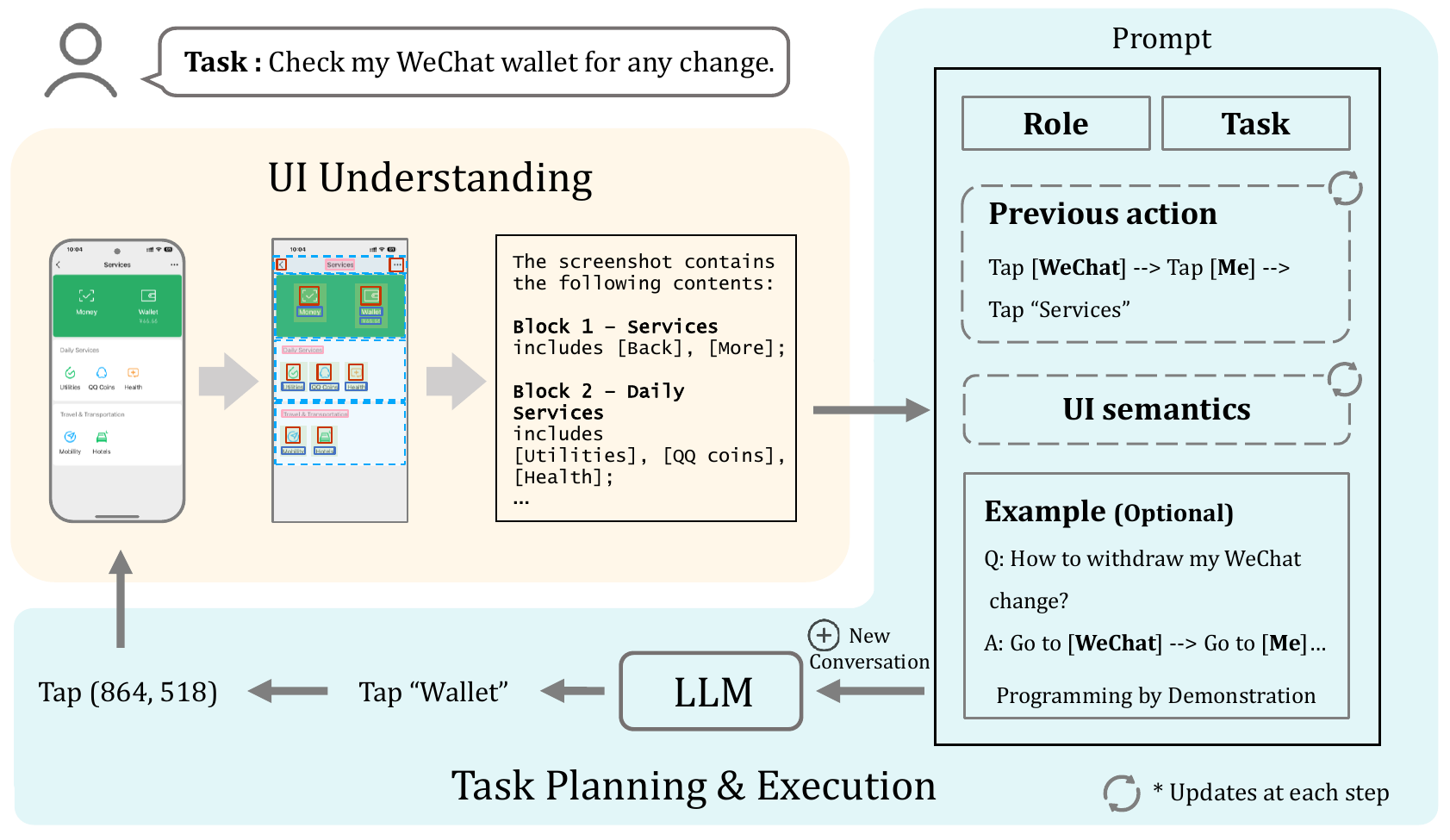}
    \caption{The workflow of VisionTasker.}
    \label{fig:framework}
    \vspace{-6mm}
\end{figure}

Our goal is to develop an intelligent agent capable of performing human-like actions to automate mobile tasks in a step-by-step manner, as shown in Fig~\ref{fig:framework}. Users will simply express their desired tasks in natural language. In response, our agent takes screenshots, comprehends the UI's current semantic context, and devises a strategic action plan. It then executes low-level interactions—such as taps and swipes—to advance the task autonomously. After executing each action, the agent reviews the new screen and previous actions to guide its subsequent move. This iterative process persists until the task is complete or some action limit is reached. Meanwhile, users are liberated from hands-on interactions, yet they can still monitor the task's progression through visible cues, like text inputs or option selections. Importantly, users retain the power to halt the process at any moment, ensuring their control over the agent. 

\begin{figure}
    \centering
    \includegraphics[width=1\linewidth]{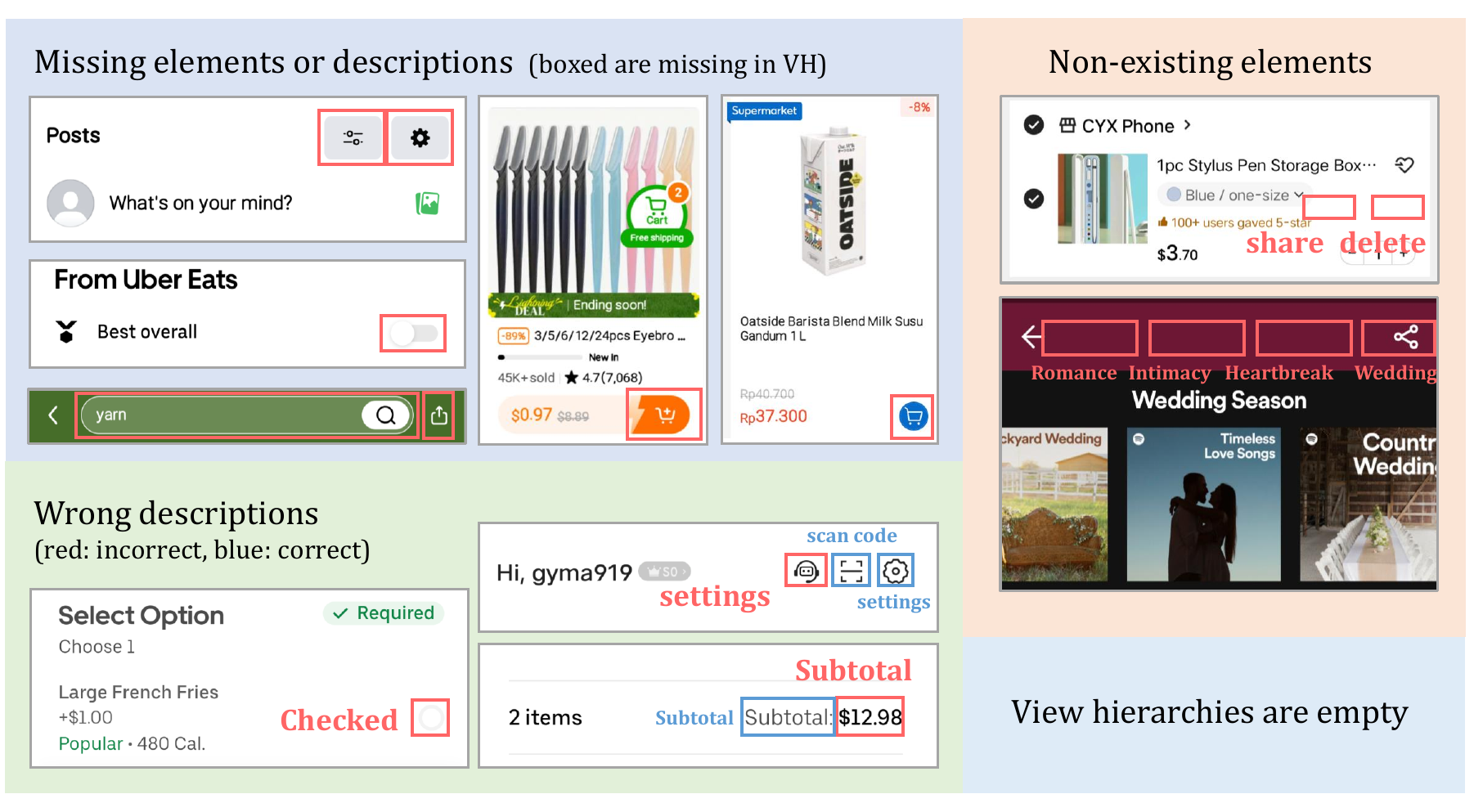}
    \caption{Cases where using view hierarchies for UI description is problematic. Screenshots from Facebook, Uber, Spotify, Temu, Shopee, and Shein, each with 100M+ installs.}
    \label{fig:vh_errors}
\end{figure}

A large portion of previous works on mobile task automation rely on view hierarchies, which are a kind of Android UI metadata containing a tree-based representation of an application's UI, as a source for the description of UI. However, view hierarchies are not always accessible and can suffer from issues like incomplete object descriptions or misaligned structural information~\cite{ross2018examining,li2022learning,XDA}, which may hinder the model's effectiveness and generalization. Figure~\ref{fig:vh_errors} shows some issues that can arise when using view hierarchies in common apps, including missing elements or descriptions, empty view hierarchies, wrong descriptions, and the inclusion of non-existing elements.

Our approach's viability hinges on the principle that a well-crafted UI, being expressive and informative, allows even those unfamiliar with it to grasp its semantics and intuitively navigate the application. This principle is based on the alignment of UI design's semantics and logic with users' mental models~\cite{nielsen1994enhancing}. In accordance with this, we simulate human comprehension of the UI by integrating the visual layout and the content of UI elements, as identified or extracted by our vision models, to output the UI's semantics. Furthermore, the recent advancements in LLMs, propelled by their extensive training on vast datasets of human-generated text and their significant size, have equipped LLMs with the ability to mimic human knowledge acquisition and reasoning processes. Leveraging this, our intelligent agent uses an LLM as a navigator for human-like task planning, applying the UI semantics decoded by vision models to translate high-level human instructions into step-by-step machine commands effectively.

The intelligent agent comprises three key components: the UI understanding module, the task planning module, and the execution module. In the UI understanding module, components like widget detectors and text recognizers analyze screenshots to identify widgets, text, and their locations. These elements are then grouped based on the visual UI layout and output in natural language, forming distinct blocks to convey varied semantics for task planning. The task planning module relies on a LLM to break down and plan tasks into steps. Taking input in the form of a prompt containing task description, action history, and, UI semantics, it generates specific low-level actions required in the current UI to advance the task, expressed in plain language, such as ``tapping the Settings button on the current UI''. The execution module interprets these text-based actions into corresponding operation commands by associating the objects with the UI elements provided by the UI understanding module and determining the coordinates of the operations, like ``sending a touch-down event at coordinates (X, Y)''.

\begin{figure*}
    \centering
    \includegraphics[width=1\linewidth]{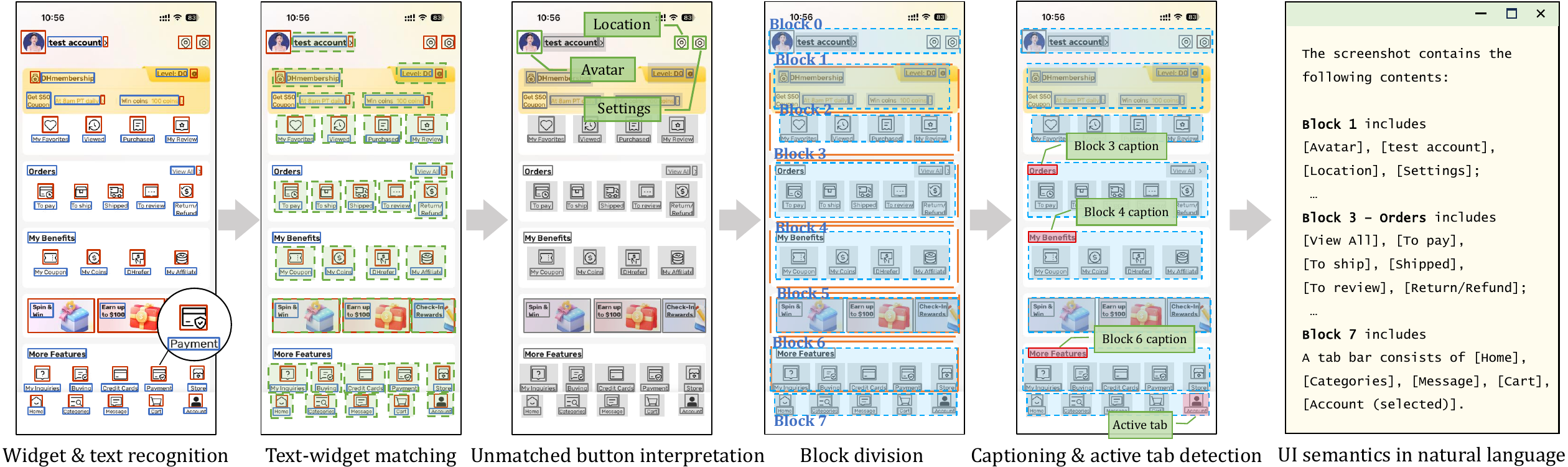}
    \caption{The process of vision based UI understanding.}
    \label{fig:ui_understanding}
\end{figure*}

\section{UI Understanding}
Unlike previous methods that depend on view hierarchy or accessibility services to access UI elements, our module directly examines UI screenshots. This approach enables us to tap into the extensive semantic data within the UI's visual structure comprehensively. It detects text, buttons, and various widgets, taking into account the dynamic and interactive features of the UI, and organizes them into intelligible, meaningful clusters. The output of this module is a detailed description of the UI semantics in natural language. The whole process is shown in Fig~\ref{fig:ui_understanding}.

\subsection{Widget Detection}
\label{sec:widget}
Understanding the UI first requires accurately identifying and recognizing its various graphic widgets. This is also essential for grasping the action space of the interface and performing actions such as tapping certain buttons or typing in text boxes. To accurately recognize the varied widgets within the UI, we carefully retrain the state-of-the-art model, YOLOv8~\cite{Jocher_YOLO_by_Ultralytics_2023}, specifically for UI object detection, leveraging open-source frameworks to optimize its performance for UI understanding. YOLOv8 employs a Convolutional Neural Network architecture and has undergone training on a substantial dataset of photos from everyday scenarios, allowing it to adeptly detect and recognize objects like apples, bikes, and buses. However, as UI significantly differ from everyday scene photos, and UI elements don't align directly with the class labels in these models, we craft a training set tailored for UI using publicly available datasets RICO~\cite{deka2017rico} and VINS~\cite{bunian2021vins}, and retrain the models. The output of the models consists of all the detected elements with their class labels and bounding box coordinates. Given the long-tail distribution of widgets in UI, at this stage, we focus on the 12 most common classes, including status bar, navigation bar, button, edit text, image, page indicator, seek bar, rating bar, check box, radio group, spinner, and switch.

Owing to the uneven distribution of samples across the 12 classes in the dataset (for example, a large number of buttons but a scarcity of text boxes), we employ a synthetic UI approach for data augmentation. In the first step, we create a canvas with a \textcolor{black}{random} color, add a status bar and a navigation bar by stretching them horizontally to fit the canvas size, and position them at the top and bottom to serve as the UI background. Then, elements are sequentially added to the canvas. For example, to address the Page Indicator, we catalogue every occurrence, assess the distribution of their placements within the UI landscape of our dataset, and proceed to randomly augment characteristics such as brightness, contrast, color balance, and transparency of a chosen Page Indicator. This modified widget is then positioned according to its observed distribution. 
\textcolor{black}{In total, we synthesize 4856 UI images using element patches from various real UI images. The augmented elements include five types that are less presented in the dataset: ratio group, page indicator, spinner, seek bar, and rating bar.} The resultant synthetic UIs are then merged into our training dataset. By injecting controlled variability into the UI generation process, this method not only compensates for the scarcity of certain elements in the original dataset but also significantly enriches the dataset's diversity, thereby broadening the training spectrum for underrepresented elements.

\subsection{Text Recognition}
In addition to graphic widget elements, user interfaces typically include rich textual information. To extract text and its location from screenshots, we employ an implementation of the open-source PaddleOCR framework\footnote{https://github.com/PaddlePaddle/PaddleOCR}, which consists of two main components: text detection and text recognition. The text detection component identifies text regions within input images using the PP-LCNetV3 model. This model is extensively trained with text images, allowing it to accurately locate text, even when it varies in terms of position, orientation, or size within the image. Meanwhile, the text recognition module is responsible for extracting specific text content from the regions identified by the detection module. It performs OCR processing on these detected text regions, utilizing the SVTR model~\cite{ijcai2022p124} to handle sequential data. Through the connectionist temporal classification loss~\cite{graves2006connectionist} and GTC optimization method~\cite{hu2020gtc}, it accomplishes character-level recognition of text regions and assembles them into complete text content. By combining these two components, we can obtain both the text and its corresponding location within the screenshot. It's important to note that due to the presence of various graphical patterns or icons in UI, which can potentially interfere with text recognition, we have set a high threshold for recognized texts. Only content recognized with a confidence value exceeding 0.95 is recorded as text.

\subsection{Semantic Grouping}
Widget detection and text recognition enable identifying the types of interactive widgets, extracting the content of the text, and determining the positions of each element. This section will explore how to integrate these elements into a cohesive and semantic whole based on the UI layout.


\subsubsection{Text-Widget Matching}
The appearances and semantics of low level graphic widgets are closely related to the high level design of the app's UI. Consequently, even widgets with the same function may exhibit different styles in varied applications, and conversely, widgets with similar styles may serve different purposes. To alleviate user confusion stemming from these variations, important widgets such as buttons in the UI are typically accompanied by explanatory text nearby, providing explanations and cues for their functions. Therefore, it is crucial to correctly align the widgets with the associated text in the interface to retrieve true semantics of various widgets.

Our approach to text-widget matching leverage an observation that when humans encounter a collection of UI elements, they tend to treat elements in close proximity as a cohesive whole. This observation forms a key aspect of UI design known as the proximity principle~\cite{wertheimer1938laws}, where items placed close together are perceived as more closely related than those spaced farther apart.
Building on this principle, we use a proximity-based method to match texts with widgets inspired by~\cite{xie2022psychologically}. For various widgets and texts identified, we utilize their bounding box coordinates to serve as representations for each element. We determine widget-text matches by calculating distances between a widget and its text neighbors, considering those with a distance below the maximum threshold. And a widget can match with one or more texts if they are aligned to maintain continuity. In certain instances, such as with app icons on the home screen, where only the widget portion is clickable, we merge widget-text matches by retaining only the widget. Despite its simplicity, this approach proves to be effective, facilitating the establishment of associations between widgets and textual elements. This, in turn, permits the leveraging of text data for precise widget interpretation, significantly improving the descriptiveness and quality of our annotations.

\subsubsection{Unmatched Button Interpretation}


Unmatched buttons refer to buttons with no matched text. They are usually graphical elements that perform specific actions or lead to certain functionalities within the app or website but lack explicit text explaining their purpose. Human users are often expected to understand these buttons' functions based on their design, context within the UI, or common conventions (e.g., expressive icons such as a magnifying glass for search, a gear for settings). For unmatched buttons, an underlying design principle is the appearances or related icons are sufficiently familiar to users, negating the need for textual explanations. To effectively interpret unmatched buttons, we utilize the capabilities of CLIP, a powerful pre-trained multi-modal model. By fine-tuning CLIP~\cite{radford2021learning}, we enhance its proficiency in inferring the functions of these buttons based on their visual design. This is particularly useful in our context as CLIP excels in understanding both images and text within a joint embedding space, enabling it to interpret various meanings of non-text-buttons in real-world UIs.

We employ CLIP to match a non-text-button with one text in a set of descriptions for common button functions such as ``Setting'', ``Send'', ``Submit'', etc. Let \(V=\{v_1,v_2,...,v_n\}\) denote the images of the buttons, and \(T=\{t_1,t_2,...t_n\}\) denote their corresponding text descriptions. Our goal is to fine-tune a CLIP model, aligning both the images and text within the same latent space. To achieve this, we utilize the IconSeer dataset~\cite{li2023you}, a publicly available resource comprising over 180k buttons with text descriptions spanning more than 170 common categories for the fine-tuning process. We employ a contrast loss function:
\[
\mathcal{L} = -\frac{1}{2N} \sum_{j=1}^{N} \left[ \log \sigma(v_j, t_{+,j}) + \log \sigma(v_{+,j}, t_j) \right]
\]
where \( v_j \) is the image embedding for the \( j \)-th pair, \( t_{+,j} \) is the matching text embedding for the \( j \)-th text, \( \sigma(a, b) \) represents the softmax term \(\frac{\exp(a \cdot b / \tau)}{\sum_{i=1}^{N} \exp(a \cdot k_i / \tau)}\). Here, \( N \) is the batch size, and \( \tau \) is a temperature scaling parameter. For rapid inference, we opt for a small model of ViT-B/32\footnote{https://huggingface.co/sentence-transformers/clip-ViT-B-32}. 

\subsubsection{Semantic Block Division}
In UI design, it's common to group similar functions and information in distinct sections, or ``blocks'', to make navigation easier and reduce the effort it takes for users to find what they need. These blocks are visually set apart from their surroundings, often using distinct colors to create clear border lines. By identifying such border lines in UIs, we can effectively pinpoint these semantic blocks. However, the varying orientations, lengths, and spacings of these lines in different designs present a challenge.

To address this, we employ a probability-based approach with a focus on gradient information for border detection, as shown in Alg.~\ref{alg:border} in Appendix. First, we quantize the image by reducing its colors to the most predominant ones, based on the distribution of colors in the UI. This step helps to reduce visual noise and makes borders between groups more obvious. Next, we apply edge detection to identify the significant edges. We only retain line segment edges by regarding them as connected groups of pixels and use the direction and strength of their gradients to find them. Then, using a probabilistic approach, we group these segments by how close and similarly oriented they are to identify the borders of groups. Finally, we identify blocks or group areas typically as rectangles formed by intersecting or nearly intersecting horizontal and vertical lines, as shown in Fig.~\ref{fig:ui_understanding}.

\subsubsection{Block Captioning and Active Tab Detection}

For a semantic block, there is usually a caption serving to clarify its logic, offering descriptions that improve accessibility, aid navigation, and highlight structure and interactivity, thereby providing important cues for UI understanding. To annotate captions, we examine the top area within the region to determine if there is left- or top-justified pure text content without an associated image. If such text exists, we consider it as the caption, providing additional context for the block's elements. This is especially useful in scenarios where captions fill in critical information gaps. For example, in the ticket booking UI, dates are usually shown using only the number of the day, while the corresponding name of the month is separately positioned at the top of the date list for each month. This design choice, while visually clean, can lead to misunderstandings for LLM navigators. They may recognize the month for the initial date described behind the month but fail to correctly associate subsequent dates with their respective months.

Furthermore, our method can be further utilized to identify the active tab in current UI layout. At first, we identifies the application's tab bar as a unique semantic block, attributable to its distinctive visual characteristics such as color and border lines. To pinpoint the active page within this tab bar, we introduce a recognition method based on color differences, focusing on the H (Hue) component of the HSV color space. This choice stems from the challenge of recognizing color intensity differences solely through specific RGB components without prior knowledge of the exact color. In contrast, the H component is effective in representing color variations as perceived by humans. By analyzing the intensity distribution of the H component in the buttons, we pinpoint the one with the most significant difference in distribution compared to others, attributing to it the additional semantic of the ``currently selected element''.

\section{LLM for Task Planning}
After extracting semantic information from the UI, this section discusses how to effectively utilize UI semantics to assist LLMs in task planning and introduces additional support mechanisms to address situations beyond the capabilities of LLMs.

\subsection{Planning Step by Step}
\label{sec:prompt}

Drawing from expansive datasets, LLMs possess a wealth of knowledge and the capacity for complex reasoning~\cite{brown2020language,chowdhery2023palm,wei2022chain}. Their proficiency in navigating multi-step tasks—like solving math problems—positions them as valuable navigators for automating processes. However, LLMs are not without shortcomings. At times, they may produce implausible content, a result of hallucination issues~\cite{maynez-etal-2020-faithfulness} where they invent features not present in the current app to fulfill a task. To mitigate these issues, we've taken cues from the idea of a chain of thought~\cite{wei2022chain} and adapted it into a chain of screens. This technique entails a sequential feedback loop, where the LLM uses the output of the previous action and the current UI to plan the subsequent step. This approach serves two crucial functions: it deepens the LLM’s grasp of the current interface, reducing the risk of confusion when confronted with similar interfaces, and it directs the model to select actions only from available UI elements.

The design of our prompt is streamlined into five essential parts for effective interaction with a LLM. Firstly, the \textbf{role} part establishes the LLM's role as an intelligent agent to enhance task automation by identifying and interacting with UI elements appropriately. The LLM is required to pick the element on the current UI to advance the task. 

Secondly, the \textbf{task description} allows users to articulate their objectives in natural language, such as ``Play 'Baby Blue' by Badfinger and add it to my favorites''. This is crucial for maintaining the focus of the LLM on the user's initial goal throughout the process.

Thirdly, the \textbf{action history} part records the sequence of actions executed by the LLM at each UI step, such as ``Tap Settings button -> Tap WLAN button ->...''. Keeping a record of action sequences serves two purposes: it helps in situations where the same interface requires multiple actions, like changing the UI language, which involves selecting the target language and tapping the save button on the same page. Without recording the action sequence, the LLM might keep selecting the target language and overlook the save button. It also allows for documenting any explored but failed paths during the automation process, preventing the LLM from getting stuck in a useless loop. 

The fourth part is \textbf{UI semantics}. Contrasting with previous approaches that transform view hierarchies into HTML formats, our method emulates human cognitive processes, utilizing natural language descriptions of the UI’s semantic blocks to enhance decision-making. This description is structured from top to bottom and left to right, detailing each semantic block's title (optional) and UI element within, along with the element's category, functions, and IDs. 

Finally, the \textbf{example} part is an optional and offers high-level task-solution pairs from the PBD mechanism or help documentations to improve the planning ability for related tasks.

When the LLM generates responses, it specifies both the element and the required action, such as ``Tap SAVE button''. If the LLM suggests an action on an element not present in the current UI, it's prompted to reassess and select from the elements that are actually available. This approach reduces errors and keeps the LLM focused on realistic tasks. After receiving a valid response, the conversation with the LLM is terminated. A new dialogue is initiated with the above parts by the agent when planning the next step. During this process, the action history records past actions rather than re-documenting the semantic description of each interface. This strategy is designed to prevent an excessive increase in the number of tokens due to too many turns in the conversation.

\subsection{Programming by Demonstration}
\label{sec:pbd}
In addition to relying on the inherent knowledge and reasoning abilities of LLMs for task planning, we have implemented a mechanism of programming by demonstration to address situations where LLMs may not be able to plan the correct path. This involves our UI understanding module, which bridges the gap between a task's semantic elements and their physical locations on a screen. This module interprets simple touchscreen actions, like a tap on the screen, as meaningful steps in a task. For example, a tap at a certain spot on the screen could be understood as hitting the ``save'' button, turning human actions into a sequence of semantic steps for the LLM to follow.

When the agent encounters difficulty in completing a specific user task within certain steps, it holds the task description and prompts the user for manual execution. The UI understanding module tracks each action on the screen and deciphers the higher-level meaning of those actions. For example, it might record a series of actions like ``tapping the WeChat icon on the home screen -> tapping the Plus button (indicating actions like More options, New, or Create) -> tapping the QR code button''. By integrating these user-shown steps into its prompts, the LLM can make more informed decisions. Unlike methods that just record and play back what the user does, our approach offers more adaptability. It comprehends actions at a more abstract level, allowing it to cope with changes in app layouts or updates that alter the interface. This adaptability ensures that the LLM remains effective even when facing app updates or screen resolution changes.

\section{Executing Module}

\textbf{Screenshots Capturing:} The initial step for the execution module is to capture screenshots for subsequent analysis. In fact, many app pages contain content that extends beyond a single screen view. To address this, the execution module first attempts to scroll to the bottom of the page or until a specified number of scrolls is reached (for pages that continuously load content as you scroll). Each scroll covers half the screen's length, and a screenshot is taken before each scroll. These screenshots are then stitched together to form a complete view of the current interface. During the stitching process, overlapping portions in the screenshots are identified and removed to reduce redundancy. This long screenshot is then sent into the UI understanding module to ensure we capture the entire content of a page, even if it spans multiple screen lengths.

\textbf{Command Parsing:} 
Next, the module receives instructions from the LLM in the form of a natural language description. These instructions are expected to consist of an action-object pair to guide the task forward, due to the specific instructions in the prompt. The actions are confined to tapping (including long press), entering specific text, swiping, and stopping, while the objects are usually corresponding UI elements or locations, such as ``Tap the Settings button'' or ``Enter pizza hut in the search bar''. The execution module examines whether the UI elements required by the LLM's output exist in the current interface, utilizing the list of UI elements from the UI understanding module. Should these necessary elements be absent, the execution module prompts the LLM for a reconsideration, based on the elements available in the current interface. In contrast, if the required elements are present, the execution module retrieves the bounding box coordinates of these elements for the subsequent step of execution. 

\textbf{Action Execution:} In executing commands such as ``Tap the Settings button'', the execution module initially identifies the touchpoint on the screen using the retrieved coordinates of the Settings button's bounding box. The center of these coordinates is calculated to pinpoint the exact location for tapping. In instances where long screenshots result in coordinates extending beyond the screen size, it becomes necessary to compute the distance to swipe down and recalculate the coordinates for the post-swipe interface. Hence, the actual operation involves a sequence of actions, such as (Swipe, [$Y_{start}$, $Y_{end}$]) followed by (Tap, [$X$, $Y$]). These steps are converted into Android Debug Bridge\footnote{https://developer.android.com/tools/adb} (ADB) commands, which are then transmitted to the smartphone to be carried out. Once executed, the command is recorded in the action history, facilitating the planning of subsequent steps. 

\begin{table*}
    \centering
    \footnotesize
    \renewcommand{\arraystretch}{1.2}
    \caption{Comparative analysis of three methods for UI Understanding.}
    \label{tab:7_1}
    \begin{tabular}{ccccccccccc}
        \toprule
        \multirow{2}{*}{Method} & \multicolumn{2}{c}{A1. Element detection} & \multirow{1}{*}{A2. Icon interpretation} &\multicolumn{6}{c}{A3. UI  questions answering} & \multirow{2}{*}{Token}\\
        \cmidrule(lr){2-3} \cmidrule(lr){4-4} \cmidrule(lr){5-10}
        & Precision& Recall &   Accuracy   & Status & Content & Count & Functionality & Summary & Overall &     \\
        \midrule               
        GPT-4V~\cite{OpenAIGPT4V2023}&    -\textsuperscript{$a$} &    -\textsuperscript{$a$} & 0.83 & \textbf{0.67} & 0.73 & 0.54 & 0.71 & 0.98 & 0.75 & 1105\textsuperscript{$b$}\\
        VH~\cite{wang2023enabling}    & 0.37 & 0.68 & 0.63 & 0.20 & 0.73 & 0.74 & 0.56 & 0.92 & 0.63 & 1267 \\
        \rowcolor{lightgray!40}
        Ours  & \textbf{0.94} & \textbf{0.94} & \textbf{0.85} & 0.57 & \textbf{0.91} & \textbf{0.88} & \textbf{0.90} & \textbf{0.98} & \textbf{0.87} & \textbf{265}  \\
        \bottomrule
        \multicolumn{10}{l}{\textsuperscript{$a$}GPT-4V cannot accurately output the specific coordinates of UI~\cite{yang2023set}, hence it was not adopted in the A1 evaluation.}\\
        \multicolumn{10}{l}{\textsuperscript{$b$}An image with dimensions of 1920x1080 consumes 1105 tokens in the GPT-4-1106-vision-preview model.}\\
    \end{tabular}
\end{table*}

\section{Evaluation}
In this section, we conducted extensive experiments to evaluate our approach in four key aspects: its accuracy in recognizing diverse UI elements and interpreting unmatched buttons, the effectiveness of our vision-based UI understanding module in enhancing accurate task planning, the approach's performance in real-world tasks, and the impact of the PBD mechanism in challenging scenarios faced by the LLM.
\subsection{Performance of UI Understanding}
\label{sec:7_1}

\aptLtoX[graphic=no,type=html]{\begin{figure}
        \centering
        \includegraphics[width=1\columnwidth]{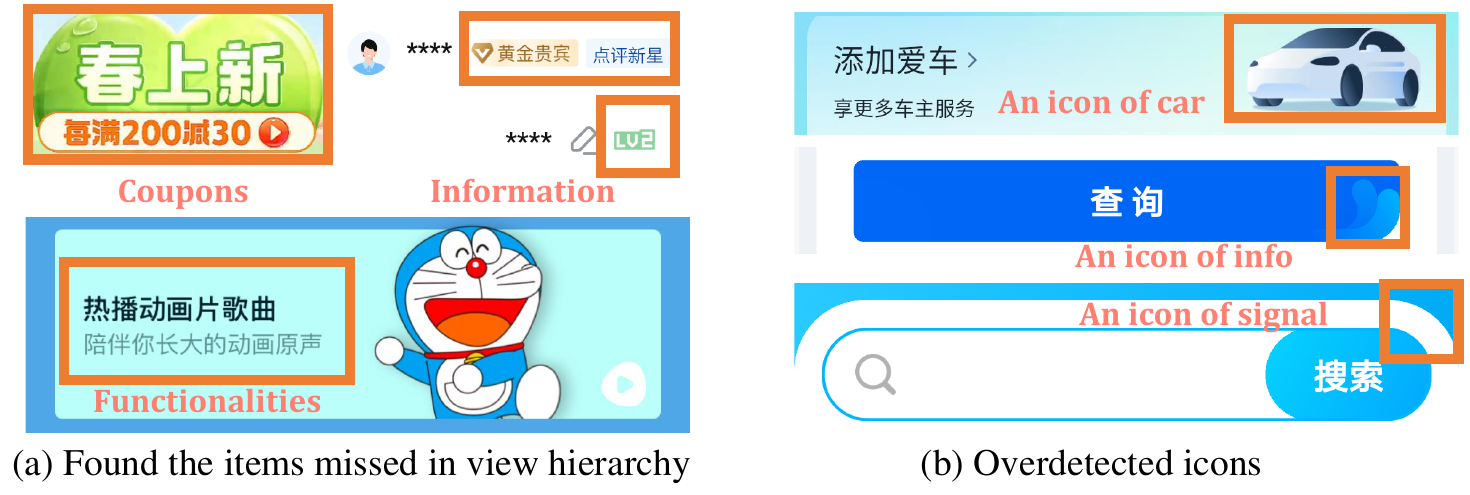}
        \caption{Our method finds more details and non-exist elements.} 
        \label{fig:image_read}
    \end{figure}
    \begin{figure}
        \centering
        \includegraphics[width=1\columnwidth]{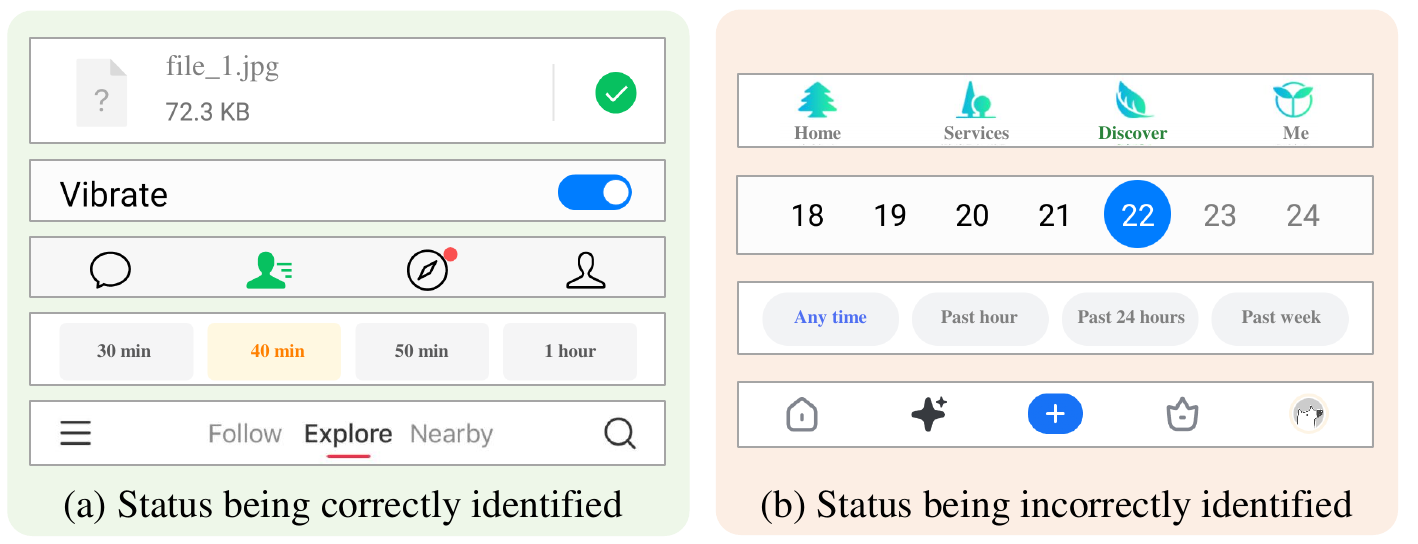}
        \caption{Cases of correctly and incorrectly identified status.}
        \label{fig:cases}
   \end{figure}}{\begin{figure*}
    \begin{minipage}[t]{0.52\linewidth}
        \centering
        \includegraphics[width=1\columnwidth]{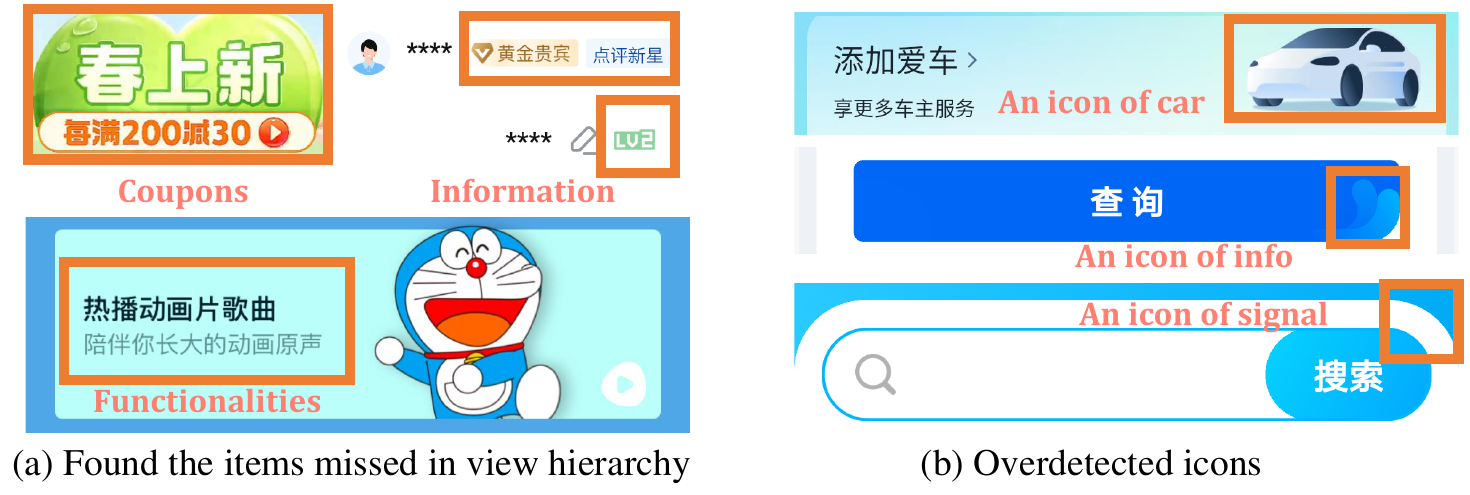}
        \caption{Our method finds more details and non-exist elements.} 
        \label{fig:image_read}
    \end{minipage}
    \begin{minipage}[t]{0.47\linewidth}
        \centering
        \includegraphics[width=1\columnwidth]{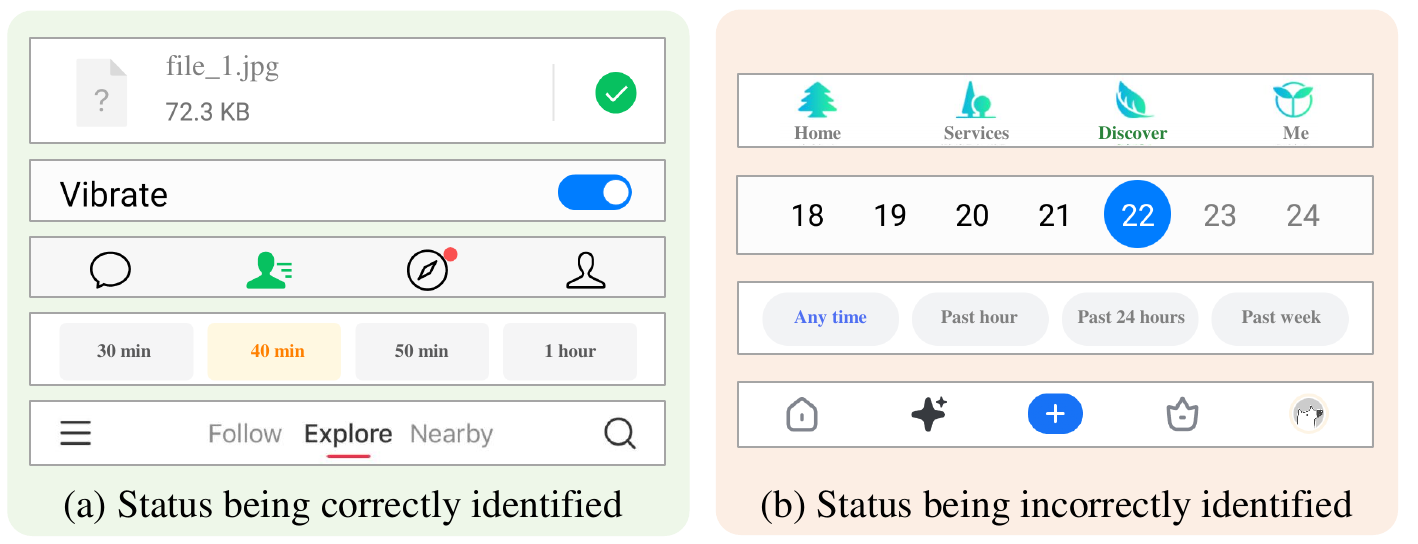}
        \caption{Cases of correctly and incorrectly identified status.}
        \label{fig:cases}
    \end{minipage}
\end{figure*}}

Our UI understanding method can analyze screenshots to identify the positions and functions of UI elements and provide a natural language description of the entire screenshot. We assess the performance of our method through three key dimensions: A1. accuracy in detecting UI elements, A2. the ability to interpret the meanings of unmatched icon-only buttons, and A3. the effectiveness of generated screen semantic descriptions for LLM to answer UI related questions. 

\subsubsection{Dataset}
Our evaluation employs a dataset of 136 screens from 31 widely-used applications, including 16 built-in system apps and 15 top apps from the app store, each with over a billion installs. Our selection aimed to capture a diverse array of layout patterns to enhance the diversity of interfaces, as depicted in Fig.~\ref{fig:layout} in Appendix. To ensure our model's broad applicability and to prevent any overlap between training and testing datasets, we trained our vision-based detection and interpretation models on UI screenshots from English apps collected years ago from the Google App Store. In contrast, for testing, we used the most recent Chinese apps in the Huawei App Store. 

Three authors carefully annotated the selected screens, including identifying all textual and graphical elements by marking their bounding box coordinates (A1), annotating the function description of icon-only buttons (A2).

For UI question answering by LLM (A3), we design a series of question-and-answer pairs for each UI screen. Example questions are shown in Table~\ref{tab:question_exp} in Appendix. The questions are crafted examining UIs from four perspectives: 
\begin{itemize}
    \item \textit{Status inquiry}: questions about the state of elements, e.g., the on or off status of the VPN.
    \item \textit{Content questions}: inquiries regarding the text information displayed in text or images, e.g., the customer service numbers shown in the UI.
    \item \textit{Item count}: requests for counting specific items displayed, e.g., the number of flights shown on a screen.
    \item \textit{Functionality query}: questions concerning the execution of specific function within the UI, e.g., how to adjust the number of rooms in a booking.
    \item \textit{Purpose summary}: questions aimed at summarizing the main function or purpose of the page based on its content.
\end{itemize}

Through manual annotation, we marked 4464 bounding boxes pinpointing element locations, 161 descriptive annotations for icon-only buttons, and 586 question-answer pairs. These pairs include 89 about status, 188 about content, 105 regarding functionality, 68 on item counts, and 136 on summaries.

\subsubsection{Metrics}
To evaluate the performance of our element detection model (A1), we utilized the common metrics of precision and recall at an IoU of 0.5 from the object detection domain~\cite{bunian2021vins}. Precision reflects the proportion of predicted bounding boxes that accurately match manually annotated boxes, while recall measures the extent to which true bounding boxes are successfully predicted by the model. For A2 and A3, accuracy is used as the metric. Three authors independently reviewed the model generated answers and reached a consensus to resolve any discrepancies. 


\subsubsection{Baselines}
We set two baselines for comparison, incorporating both a vision-based approach (GPT-4V~\cite{OpenAIGPT4V2023}) and a view hierarchy (VH) based method ~\cite{wang2023enabling}. GPT-4V is the state-of-the-art image-to-text model and exhibits remarkable visual comprehension and inference capabilities. Unlike two-stage processes that separate UI understanding from task planning, GPT-4V processes screenshots and instructions together as input, allowing it to respond to queries directly. However, its limitation in generating precise locations for interface elements precluded its use in assessing A1. The VH based approach detailed by~\cite{wang2023enabling} presents a novel technique of extracting useful information from the view hierarchy, translating it into HTML format as UI description. This method directly accesses element positions within the view hierarchy, and is shown effective in providing a comprehensive UI description for LLM understanding. For the question-answering task (A3), we utilize an Ernie Bot to answer four types of questions based on generated UI semantic descriptions  (VH in HTML format and ours in natural language). The results are shown in Table~\ref{tab:7_1}.

\subsubsection{Results and Analysis}


Our method for detecting UI elements (A1) significantly surpasses the accuracy of VH-based approaches. The inherent limitation of VH is its frequent failure to accurately represent UI elements, often omitting existing elements or including non-existent ones. In contrast, \textcolor{black}{our visual-based method effectively identifies UI components even when applied to apps from different cultures, such as training on English apps and testing on Chinese apps. Examples are shown in Fig.~\ref{fig:image_read}(a). However, it is important to note that there is a minor possibility of false positive detections, where the method may occasionally identify non-existent icons, as shown in Fig.~\ref{fig:image_read}(b).}

The challenge extends to recognizing icon-only buttons (A2), where VH-based methods only manage to accurately describe 63\% of such elements. A contributing factor to this shortfall is that programmers often fail to provide clear descriptions for common icons or resort to using app-wide custom encodings (such as private Unicode characters), which VH struggles to interpret correctly. On the other hand, our method, along with GPT-4V, demonstrates superior performance in recognizing these commonly used icons, achieving an accuracy rate exceeding 80\%.

For LLM question answering using generated UI semantic descriptions (A3), our method outperforms the other two approaches overall. It's noteworthy that, for queries about element status, GPT-4V leads in accuracy due to its advanced image understanding capabilities. Our analysis reveals a significant gap in VH files regarding the annotation of element states (such as on/off, selected/unselected), a detail often overlooked by programmers in XML files or code. Our method adeptly recognizes the status of elements through additional icons such as toggle buttons and checkboxes. When it comes to identifying selected items in tabs or navigation bars, our method relies on comparing the main color differences between elements. While this approach is effective for simpler designs, it struggles with more complex layouts in tabs or navigation bars. Fig~\ref{fig:cases} shows the correct and incorrect cases of our method in recognizing the status.

For item count related queries (A3), such as listing all the orders or flights displayed, our method also achieve better accuracy than the baselines. This is attributed to our approach of grouping UI elements based on their layout and visual cues, enabling to more accurately aggregate related items in natural language descriptions. However, GPT-4V falls short in this area due to two identified issues: the generation of hallucinated items that don't exist and an incomplete understanding of the UI, often only recognizing items arranged in the first row under the target description and disregarding subsequent lines as unrelated.

For content related questions (A3), our method has advantages of dealing with images that lack text descriptions by utilizing OCR to read key information in images, such as product descriptions in showcase images or tracking numbers in photos of shipping labels. Similarly, for functionality queries, many functions are executed through image buttons, which might not be accurately described in the view hierarchy. Thus, our approach surpasses VH-based methods by providing a richer understanding of the UI information and functionalities. 


\subsection{Performance on One-step Prediction across Public Datasets}
\label{sec:7_2}

\begin{figure}
    \centering
    \includegraphics[width=1\columnwidth]{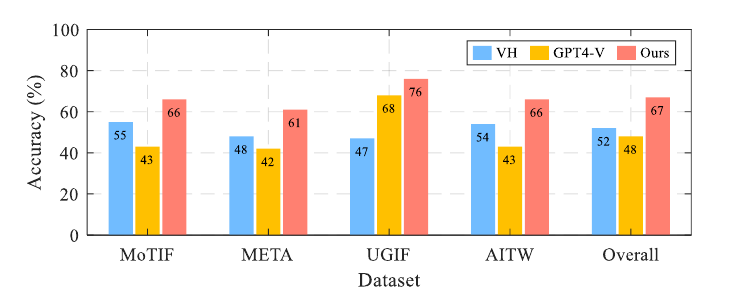}
    \caption{\textcolor{black}{Accuracy of one-step prediction across four datasets.}} 
    \label{fig:one_step_new} 
\end{figure}

\begin{table}
    \footnotesize
    \caption{Error analysis of the one-step prediction experiment.}
    \label{tab:7_2}
    \begin{tabular}{cccccc}
        \toprule
        \multirow{2}{*}{Method} & \multicolumn{3}{c}{UI error} & \multirow{2}{*}{\makecell{Planning\\error}} & \multirow{2}{*}{\makecell{Total\\error}}\\
        \cmidrule(lr){2-4} 
        & Missed & Overdetected & Misread   &  &   \\
        \midrule               
        GPT-4V~\cite{yan2023gpt}& 12 & 2 & -\textsuperscript{$c$} & 232 & 246\\
        VH~\cite{wang2023enabling} & 27 & 8 & 37 & 154 & 226 \\
        \rowcolor{lightgray!40}
        Ours  & 12 & 5 & 12 & 126 & 155 \\
        \bottomrule
        \multicolumn{6}{l}{\textsuperscript{$c$}GPT-4V inputs the screenshot with tags to concurrently analyze the elements} \\ 
        \multicolumn{6}{l}{and plan the next step, so the misread and planning errors are combined. }\\
    \end{tabular} 
    \label{tab:error}
\end{table}

This section investigates whether our generated natural language descriptions of UIs support LLMs to plan tasks effectively.

\subsubsection{Datasets}
To explore this, we carried out a comparative analysis, setting our method against two recent methods across four expansive UI control benchmark datasets: MoTIF~\cite{burns2021mobile}, META~\cite{sun2022meta}, UGIF~\cite{venkatesh2022ugif}, and AITW~\cite{rawles2023androidinthewild}. These datasets document human-device interactions across different tasks, including a series of screenshots, view hierarchies, user actions for each task. Due to the lack of real-time feedback following action execution in the datasets, in this section, we focus on just predicting the immediate next step based on the current task and interface. Following the experiment settings in~\cite{rawles2023androidinthewild}, \textcolor{black}{we randomly selected 470 distinct tasks from these datasets. From the action sequences to complete each task, we sampled a snapshot to represent the current state for predicting the subsequent action to accomplish the task. The testing samples included 98 tasks from MoTIF, 95 from META, 93 from UGIF, and 184 from AITW.}

\subsubsection{Baselines}
Similarly, we selected the VH based approach (VH + LLM)~\cite{wang2023enabling} and a GPT-4V~\cite{yan2023gpt} based methods as the baselines. To highlight the performance differences resulting from UI semantics, we employed ERNIE Bot as the LLM coordinator for both VH and our method. We unified the prompt across the two methods to maintain consistency. The prompt includes instructions for the LLM (``Supposing you are an intelligent agent to help users complete mobile tasks. Given the screen, predict the element in the current UI to complete the task''), a task description, and UI semantics (HTML representations for VH or natural language descriptions for ours). We intentionally omitted action history from the prompt to compel the LLM to base its predictions solely on the current UI semantics.

GPT-4V's capability to directly process images and instructions allows it to bypass the need for a LLM coordinator. However, its limitation in outputting precise coordinates prevents direct comparison with the dataset's ground truth to evaluate the accuracy of its responses. To address this, we adopted the approach in~\cite{yang2023set}, which involves detecting UI elements and their bounding boxes first using our approach. We then identified these elements by placing tags at the center of their bounding boxes. GPT-4V was tasked with predicting the tags of the elements to interact with, allowing to parse the predicted interaction locations into accurate coordinates to compare with the ground truth. For these experiments, we used the latest gpt-4-vision-preview\footnote{https://platform.openai.com/docs/models/gpt-4-and-gpt-4-turbo} API and the prompt is similar to the other two methods, except it did not include the UI semantics descriptions in text.

\subsubsection{Results and Analysis}
The results presented in Fig.~\ref{fig:one_step_new} demonstrate the general effectiveness of various methods in predicting the next step based on the current UI, demonstrating the efficiency of these UI representations. In over half of the cases, these methods were able to accurately forecast the subsequent step. Notably, our approach outperformed the others, \textcolor{black}{achieving an accuracy rate of 67\% across four datasets—a substantial improvement of over 15\% compared to baseline methods.} Upon closer examination of the errors (see Table~\ref{tab:error}), we discovered they primarily fall into two categories: errors in understanding the UI, where the model fails to correctly identify the UI elements essential for completing the task, and errors in task planning, where, despite recognizing necessary elements, the model does not correctly plan the next move.

Delving deeper into the UI understanding errors, our data categorizes them into three specific types: missed detections, where the ground-truth elements that should have been identified were overlooked; overdetections, where non-existent elements were incorrectly recognized as part of the next step; and misread, where elements were correctly detected but their functions or statuses were misunderstood, leading to erroneous interactions. Our method significantly reduced these types of errors compared to the VH method, suggesting that visual-based approaches have the potential to overcome the limitations of view hierarchies, including information omission, false positives, and inaccuracies.

Beyond UI misunderstandings, more errors arose during the task planning phase. The semantic descriptions of interfaces may fail to support LLMs in making accurate plans, leading to incorrect predictions of the element to interact with—even when both the element and the ground truth were correctly understood within the UI descriptions. Our approach resulted in 21 fewer errors compared to HTML-based UI descriptions used in the VH method. This improvement suggests that our natural language description method, which utilizes semantic blocks to enrich element semantics, provides more effective support for LLMs in task planning. This advancement not only highlights our method's superiority in understanding UIs but also in facilitating more accurate task execution by AI models.

\begin{figure}
    \centering
    \includegraphics[width=1\linewidth]{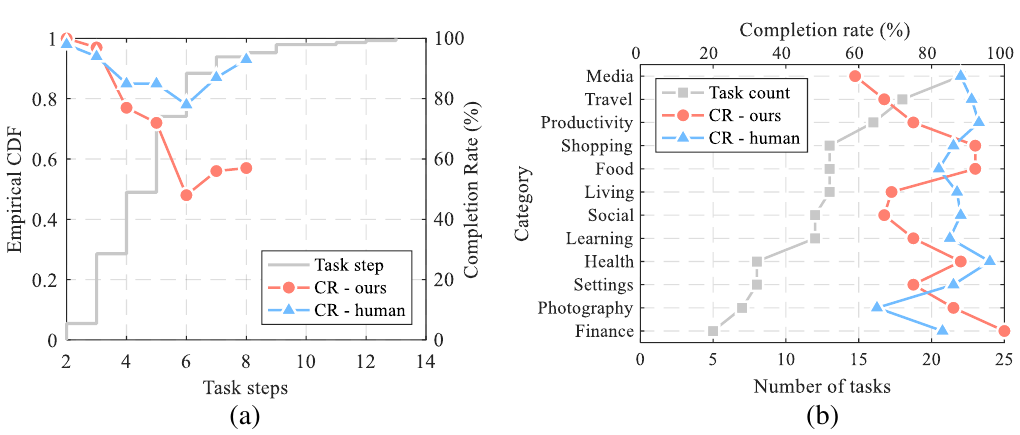}
    \caption{\textcolor{black}{Steps and categories with task completion rates.}}
    \label{fig:category}
\end{figure}

\begin{figure*}
    \centering
    \includegraphics[width=1\linewidth]{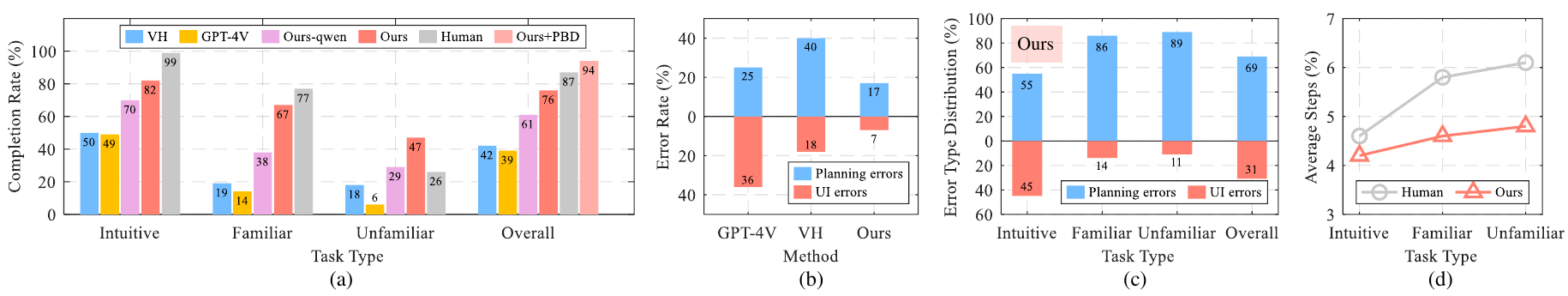}
    \caption{\textcolor{black}{Results of the real-world tasks automation experiments. ``Ours-qwen'' refers to the implementation of our framework using the open-source Qwen while ``Ours'' indicates the use of the Ernie Bot as the LLM."}}
    \label{fig:task}
\end{figure*}

\subsection{Real-world Task Automation Experiment}
\label{sec:7_3}
In addition to conducting simulated experiments on public datasets, in this section, we will explore the performance of our method in completing real-world multi-step tasks.

\subsubsection{Dataset and device}
We designed 147 real-world multi-step tasks to test our approach on a HUAWEI P20 smartphone. The device runs on the Android operating system with a 5.8-inch screen and a resolution of 2244$\times$1080. It was connected to a desktop running a smart agent through ADB, enabling the transmission of interface screenshots to the agent and the execution of commands sent by the agent. The tasks spanned 12 categories (see Fig~\ref{fig:category}(b)) across 42 commonly used apps in China and required between 2 to 13 steps (with an average of 4.7, the distribution of these task steps is illustrated in Fig~\ref{fig:category}(a)) to complete. All the tasks are listed in Appendix~\ref{sec:tasks}. A task was considered complete when the intended goal was achieved and the agent recognized the appropriate point to cease actions. We used the task completion rate as a key metric for evaluating performance. 

\subsubsection{Baselines}
\textcolor{black}{For comparison, we recruited 12 human evaluators to manually perform these 147 tasks with the same device as the human baseline. These evaluators were from local institutions, aged between 22 and 35}, with at least a bachelor's degree and over two hours of daily smartphone usage, thus proficient in various everyday mobile tasks. They were shown the descriptions of all the tasks and instructed to complete the tasks in as few steps as possible, with the option to skip any step they didn't know how to complete. One author supervised the user study for assistance. All apps related to the tasks were pre-logged in, and no personal information was needed for the study. The study lasted approximately 70 minutes, and evaluators were free to stop at any time. They were compensated based on the local average hourly wage. \textcolor{black}{We also include two additional baselines: the VH-based approach~\cite{wang2023enabling} and the GPT-4V method~\cite{yan2023gpt}. To further evaluate the adaptability of our approach, we substitute the Ernie Bot in our framework with the open-source Qwen1.5-110b model\footnote{https://github.com/QwenLM/Qwen2} (4-bit quantization version). This substitution serves as an additional method for comparison.}

\subsubsection{Results and Analysis}
\textcolor{black}{We grouped tasks into three categories based on how many evaluators completed them: unfamiliar (12\%, 17 tasks) with 5 or fewer evaluators completing; familiar (14\%, 21 tasks) with 6-10 evaluators completing; and intuitive (74\%, 109 tasks) with more than 10 evaluators completing. Fig~\ref{fig:task}(a) shows the completion rates for both humans and our approach. Our method completed 82\% of intuitive tasks and 67\% of familiar tasks, slightly below human rates but far above the other three methods. For unfamiliar tasks, where humans only completed 26\%, our method achieved a 47\% completion rate, demonstrating an advantage over human evaluators. Overall, our approach achieved a 76\% completion rate across all 147 tasks, close to the human evaluators' 87\%. When we added programming by demonstration prompts, our method's completion rate improved to 94\%, surpassing human evaluators. Notably, our method achieved good results using the open-source Qwen model as the LLM, significantly outperforming VH and GPT-4V. This shows that open-source models can adapt well to our method, opening up possibilities for more cost-effective, reliable deployment using open-source models.}

Our method's completion rate, as depicted in Fig.~\ref{fig:category}, varies with the number of steps and the category of tasks. Tasks with more than eight steps are grouped together due to the small sample size. There is a general decline in task completion rate as the number of steps increases. However, tasks requiring more than six steps still maintain a completion rate of around 50\%. Fig.~\ref{fig:category}(b) demonstrates that our method performs well across different task categories, with completion rates exceeding 60\%. The lowest performance was observed in media-related tasks, where two tasks failed due to the complexity of UI backgrounds (e.g., recommended media or music promotion covers), which interfered with the recognition of transparent function buttons on the UI. The other seven failed tasks involved four Hard tasks, including two tasks that all human annotators failed to complete. These tasks, featuring functions that are difficult to discover, also posed significant challenges to LLM task planning.

Similar to Sec.~\ref{sec:7_2}, we classified the errors from the methods across the tasks that were not completed into two types: UI understanding errors and task planning errors. The distribution of these error types across different methods is depicted in Fig.~\ref{fig:task}(b). The results indicate that GPT-4V encountered the highest number of UI issues, primarily due to significant hallucination problems during the experiments. GPT-4V tended to invent non-existent buttons, often required for later stages of a task but absent from the current interface. For instance, in a task involving checking the balance in a wallet, GPT-4V would instruct to click on a ``My Wallet'' button—a button necessary in the penultimate step of the task but not present on the current screen. These hallucinations likely stem from GPT-4V's extensive general knowledge, which includes similar information, yet it fails to accurately comprehend all elements on the interface, leading to such errors.

In contrast, the VH method exhibited fewer UI errors, even though one app's view hierarchy lacked all UI elements information in this study. However, this method faced the most task planning issues. This suggests that directly converting VH to HTML for UI description, as proposed in \cite{wang2023enabling}, might not perform well against the complexity of real-world interfaces. The HTML descriptions, by stacking elements sequentially, lack explicit representations of the relationships between elements. Consequently, even if the elements are correctly identified, forming a coherent semantic description to support accurate LLM decision-making is challenging. On the other hand, our method, even when employing the same LLM, can leverage the associations between text and icons, as well as the layout of the UI, to create a more accurate representation of the UI. This enhanced semantic understanding significantly supports LLM decision-making, leading to fewer task planning errors.

\textcolor{black}{
When examining how task familiarity affects our method's performance, we found that for intuitive tasks, the LLM's built-in knowledge and reasoning skills usually led to accurate task planning. Errors mainly occurred due to the agent's incomplete understanding of the UI. As tasks became less common, planning errors increased, similar to how humans struggle with planning unfamiliar tasks. In Fig~\ref{fig:task}(d), we compared the steps taken on tasks completed by both our method and at least one evaluator. The LLM often found more direct ways to complete tasks, while human users relied more on trial and error, especially for unfamiliar tasks where they made significantly more attempts. These findings suggest that the LLM can potentially be more efficient than humans, highlighting the importance of adding external knowledge to prompts to further improve the LLM's performance.
}

\begin{figure}
    \centering
    \includegraphics[width=1\linewidth]{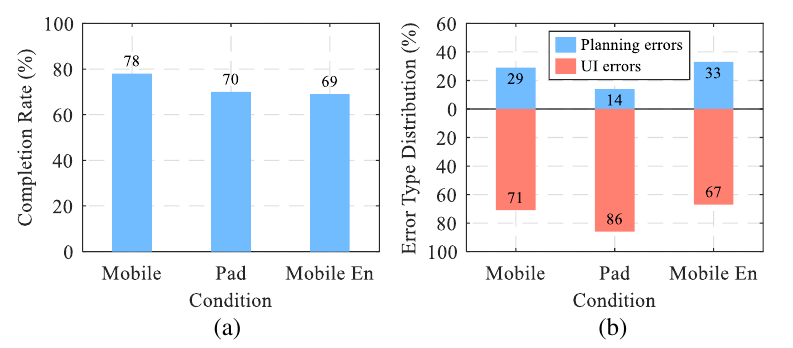}
    \caption{The effectiveness of PBD under varied conditions.}
    \label{fig:pbd}
\end{figure}

\subsection{Impact of Programming By Demonstration}
Focusing on the 36 real-world tasks that our method failed to complete, we incorporated the PBD mechanism into our approach to see if it improved performance. Specifically, an author manually executed these tasks on the experimental smartphone as the demonstration, while our agent captured a series of screens and translated them into natural language action sequences. These sequences, along with the descriptions of the tasks, were used to create 36 task-solution pairs as the PBD solutions. For each task, we enriched the prompts with the corresponding PBD solutions to examine if augmented prompts could effectively assist the LLM in accomplishing tasks that were previously unsolvable.

To rigorously test the efficacy of our PBD mechanism under varied conditions, we conducted task completion in three distinct environments: (1) on the same smartphone used for the initial demonstrations, maintaining consistency in UI and workflow; (2) on a large screen pad, specifically a 12-inch device with a 2000$\times$1200 resolution, to challenge the method with different UI layouts, screen sizes, and resolutions; and (3) on the same phone but with applications set to the English version, which might be a bit different from the original version in both layout and workflow, introducing version variation to assess adaptability to different linguistic contexts.

The results, depicted in Fig.~\ref{fig:pbd}(a), indicate that incorporating PBD significantly improves task completion, successfully automating about 70\% of the previously failed tasks. Notably, the generated PBD sequences were robust against variations in screen size, resolution, and language. As indicated in Fig.~\ref{fig:pbd}(b), our results uncovered that the primary challenge in tasks where PBD failed shifted from task planning to UI understanding. This shift highlights PBD's effectiveness in assisting with the planning of accurate task paths that LLMs find difficult. It's important to note, though, that even with PBD guidance, LLMs may still inaccurately plan tasks.

\section{Discussion}

\subsection{Latency and Token Cost}

During the task automation with VisionTasker, delays predominantly arise from two main sources: the time taken by the UI understanding module to analyze screenshots and the communication latency with the LLM. During our real-device experiments, utilizing an Intel i9 10850k CPU and an RTX 2080 Ti GPU, the UI understanding process averaged 5.9 seconds, while communication with the LLM averaged 4.5 seconds, resulting in an average execution time of over 12 seconds per action. Significantly reducing the UI understanding delays could be achieved by employing lighter models or using knowledge distillation techniques in place of the Yolov8 and CLIP models.

Another important factor is the number of tokens, which directly influences the quality of the LLM's response. In a task in Sec.~\ref{sec:7_3}, the HTML generated using the VH method consumed over 8K tokens, leading to the LLM denying service. A comparison in Table~\ref{tab:7_1} reveals that our method requires approximately a quarter of the tokens needed by the other two methods to describe the same UI. Furthermore, our prompts maintain task chains by recording completed actions as action history rather than saving previous UI descriptions, making the additional number of tokens needed due to increased task steps negligible in comparison to those needed for UI semantic descriptions. This efficiency enables the handling of tasks with longer sequences, as demonstrated by the completion of a task involving 13 steps in Sec.~\ref{sec:7_3}.

\subsection{Reading from Images}

Our method employs a visual approach to extract information from user interfaces, which might miss some elements. However, for a user-centered design, essential functionalities are generally designed to be visually prominent, often represented by standardized icons and noticeable sizes, making them more readily detectable by models. In our experiments, we observed that search boxes are the elements most likely to be overlooked by our model. This is primarily because search boxes are usually located at the top of the UI. In some applications, such as the official train ticket booking app 12306, which boasts over 1.5 billion installs, advertisements are placed in this area. These ads can make the search box semi-transparent to enhance advertising effects, thereby hindering its detection. Nonetheless, other crucial functionalities within the UI are largely detectable.

Compared to the view hierarchies that requires programmer involvement, our visual approach can potentially uncover more information but also risks detecting non-existent elements. The impact of these two types of errors on the completion of the final task differs, as the result of UI understanding serves the task planning. For instance, elements ignored by the view hierarchy, such as coupons, functionalities, and information within image buttons, which may be important to the task, can be detected by our method (Fig.~\ref{fig:image_read}(a)). The non-existent elements our method detects are mainly identified as generic images or some specifically meaningful icons. Since the wrongly identified icons' meanings are generally random and do not align with the overall functionality of the UI (Fig.~\ref{fig:image_read}(b)), these additional pieces of information, even if included in the semantic description of the UI, often do not affect subsequent tasks. Hence, using vision-based method is a viable approach for task automation in practice.

\subsection{Adapting to Desktop Tasks}
Our visual-based UI understanding method can be effectively adapted to desktop environments. Desktop UI has two distinct characteristics compared to mobile UI. First, icons in desktop apps are generally smaller; for example, a large paste button in Microsoft Word measures approximately 40x40 pixels, whereas a typical button size in mobile UI is around 150x150 pixels. This size difference means the detection model trained on mobile interfaces might not perform well in detecting desktop buttons. Second, desktop applications use less text and more icon-only buttons, relying more on the graphical representation of a button to convey its functionality, which demands higher accuracy from button interpretation models. However, these challenges can be addressed by training models with desktop interface and element datasets, without needing to alter the method's process and framework, making the adaptation to desktop environments relatively straightforward. The discussion above applies to general apps commonly used in everyday life. For highly specialized desktop apps, such as Photoshop, which feature buttons with professional meanings and where a LLM might not possess the necessary knowledge of professional workflows, our method may not yet be effectively adapted to such highly specialized apps.

\subsection{Utilizing Other LLMs}
In the task planning phase, the core competency comes from the contextual reasoning abilities LLMs acquire during their pre-training from extensive corpuses. Thus, beyond utilizing Ernie Bot, employing other LLMs like ChatGPT, Claude, and Doubao could achieve comparable or even superior reasoning outcomes. This paper primarily opts for Ernie Bot due to its ease of access, low latency, and cost-effectiveness. With the ever-improving inferencing capabilities of these models, using LLMs for task planning is becoming increasingly accurate. Our approach can also be implemented using locally deployed open-source large models \textcolor{black}{(see Sec.~\ref{sec:7_3})}, such as Qwen, LLaMA~\cite{touvron2023llama}, and ChatGLM~\cite{du2022glm}. However, achieving optimal outcomes with these models may require fine-tuning with specific datasets, such as AiTW~\cite{rawles2023androidinthewild}. Notably, our PBD mechanism can effectively apply to localized models. With sufficient demonstrations obtained through crowdsourcing or crawling from help documentations, relying on our visual UI understanding approach combined with extensive PBDs can compensate for the deficiencies of open-source LLMs in many scenarios.

\subsection{Limitations and Future Work}
VisionTasker has several limitations. For instance, some interfaces, like login screens and payment QR code scans, could not be captured via screenshots. Additionally, \textcolor{black}{accurately understanding UI remains challenging. In some cases, even when our method correctly identifies and segments all elements, it fails to fully convey the information. For example, in a local ticketing app we tested, non-bookable dates were displayed in lighter colors—a subtle distinction our method couldn't detect. Our UI understanding part doesn't incorporate advanced screen parsing methods such as screen parser~\cite{wu2021screen} and fully utilize context and graphic cues for better comprehension. VisionTasker's current approach of capturing screenshots at fixed intervals after touch events isn't optimal for games or slow-loading applications. It also isn't designed for interactive use. When a task lacks specific parameters, the LLM uses default values to complete it. For instance, in a food delivery task, orders are sent to a default address. For critical parameters, an interactive approach that prompts the user would be more appropriate. To improve our method, one could investigate advanced planning strategies such as self-reflection~\cite{shinn2024reflexion} and staged planning~\cite{li2023zero}. Testing VisionTasker on a wider range of tasks and conducting more detailed ablation studies would also help better understand and refine our method.}

\section{Conclusion}

This paper introduced VisionTasker and examined its efficacy for mobile task automation, addressing limitations found in traditional methods such as Programming By Demonstration and the challenges posed by view hierarchies. VisionTasker is empowered by a vision-based UI understanding technique, which translates UI screenshots into natural language, eliminating the need for view hierarchies required in other approaches. It integrates Large Language Models (LLM) for task planning in a step-by-step manner to enhance accuracy and practicality. Extensive experiments show that VisionTasker outperforms previous methods, providing effective UI representations across four datasets. Additionally, in automating 147 real-world tasks on an Android smartphone, VisionTasker surpasses human performance in complex tasks and showing significant improvements when integrated with the PBD mechanism, suggesting a promising direction in task automation.

\begin{acks}
We are grateful to Professor Xiaohong Guan for his kind support of this work and anonymous reviewers for their insightful comments. This work is supported by the National Natural Science Foundation of China (No. 62102308), Initiative Postdocs Supporting Program (No. BX2021243), and China Postdoctoral Science Foundation (No. 2021M702627).
\end{acks}

\bibliographystyle{ACM-Reference-Format}
\bibliography{sample-base}

\appendix

\section{Example Demonstration}
\begin{figure*}
    \centering
    \includegraphics[width=0.8\linewidth]{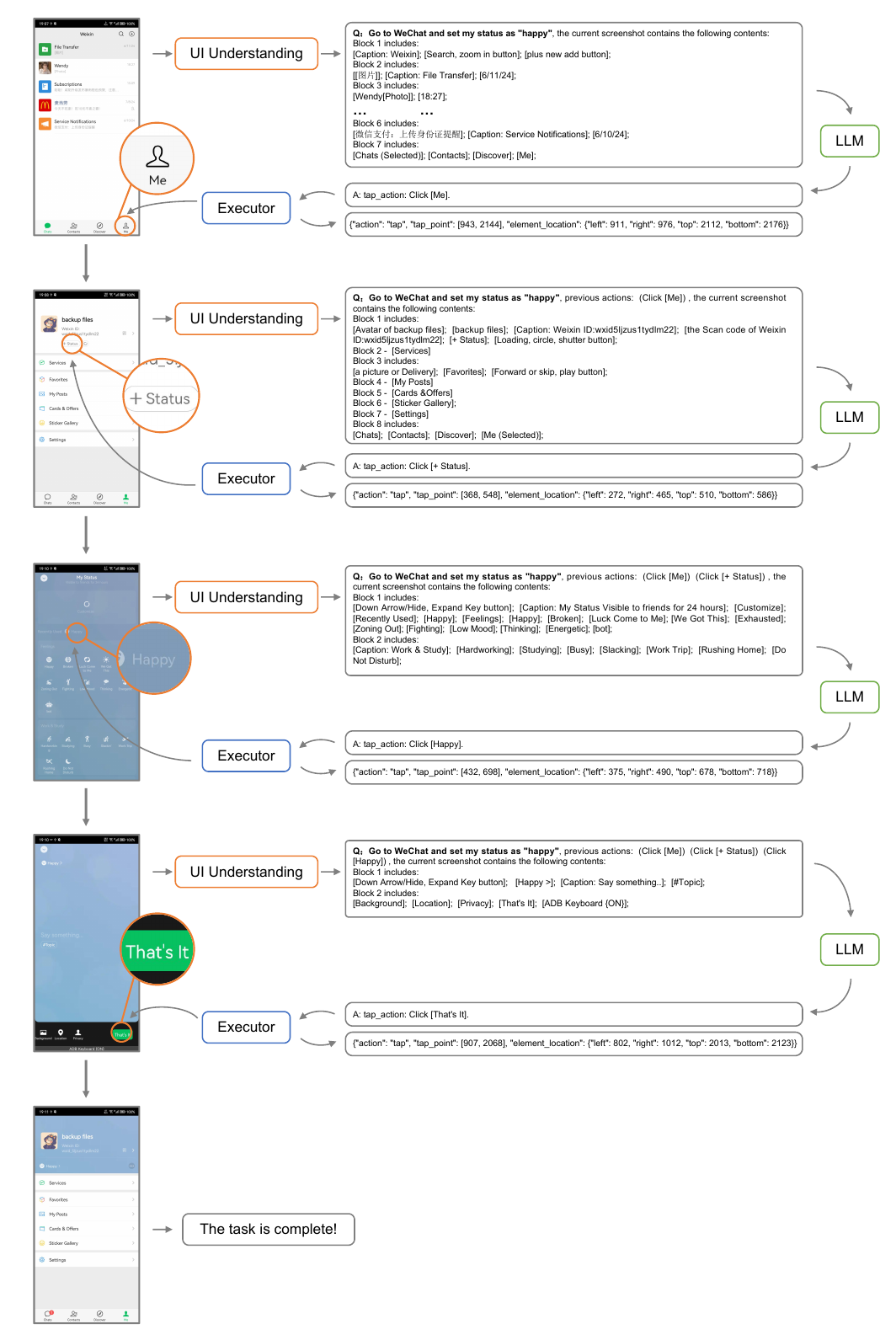}
    \caption{\color{black}{An example of how our approach can be used to complete the task step-by-step.}}
    \label{fig:whole_process}
\end{figure*}

\textcolor{black}{Figure~\ref{fig:whole_process} illustrates the demonstration of our approach in completing a sample task. The task is to set the WeChat status to ``Happy''. At each step, our method's UI understanding module converts the user interface into semantically meaningful natural language descriptions. Combining the UI description with historical actions and other information, the LLM plans and executes one step based on the current state. The plan is dynamically adjusted according to the current UI context until the task is successfully completed.}

\section{Algorithm for Block Division}
Algorithm~\ref{alg:border} shows the probability-based approach with a focus on gradient information for border detection. 

\begin{algorithm}
\small
\SetAlgoLined
\KwResult{Set of coordinate of semantic blocks $\mathcal{B}$}
    initialization: $\mathcal{L} \leftarrow \emptyset$, $\mathcal{B} \leftarrow \emptyset$;
    
    \% Convert image to grayscale\;
    $I_{gray} \leftarrow \text{ConvertToGrayscale}(I_{rgb})$\; 
    \% Identify predominant colors using a histogram\;
    $C_{topk} \leftarrow \text{FindTopColors}(I_{gray}, k)$\;
    \% Quantize the grayscale image with predominant colors\;
    $I \leftarrow \text{QuantizeImage}(I_{gray}, C_{topk})$\;
    
    Define gradient magnitude threshold $T_{grad}$\;
    \While{not end of image}{
        \% Compute gradient magnitude and orientation\;
        $M(x, y) \leftarrow \sqrt{(\partial_x I)^2 + (\partial_y I)^2}$, $\Theta(x, y) \leftarrow \arctan(\partial_y I, \partial_x I)$\; 
        \% Apply a Gaussian filter to M(x, y)\;
        $M_{filtered}(x, y) \leftarrow G(x, y) * M(x, y)$\;
        \% Edge detection using Canny detector\;
        $E \leftarrow {Canny}(M_{filtered})$\;
        \For{each edge pixel $(x, y)$ in $E$}{
            \If{$M(x, y) > T_{grad}$}{
                \% Find set of aligned neighboring pixels\;
                $\mathcal{N}(x, y) \leftarrow {AlignNeighbors}(E, x, y)$\;
                \% Fit a line segment L to pixels in N(x, y)\;
                $L = {LeastSquaresFit}(\mathcal{N}(x, y))$\;
                \If{line segment $L$ is valid}{
                    Add $L$ to $\mathcal{L}$\;
                }
            }
        }
    }
    \% Filter and group line segments into rectangles\;
    $\mathcal{B} \leftarrow {FindRectangles}(\mathcal{L})$\;
    \caption{Semantic block division}
    \label{alg:border}
\end{algorithm}

\section{Typical UI layouts}
\begin{figure}[htbp]
    \centering
    \includegraphics[width=0.85\linewidth]{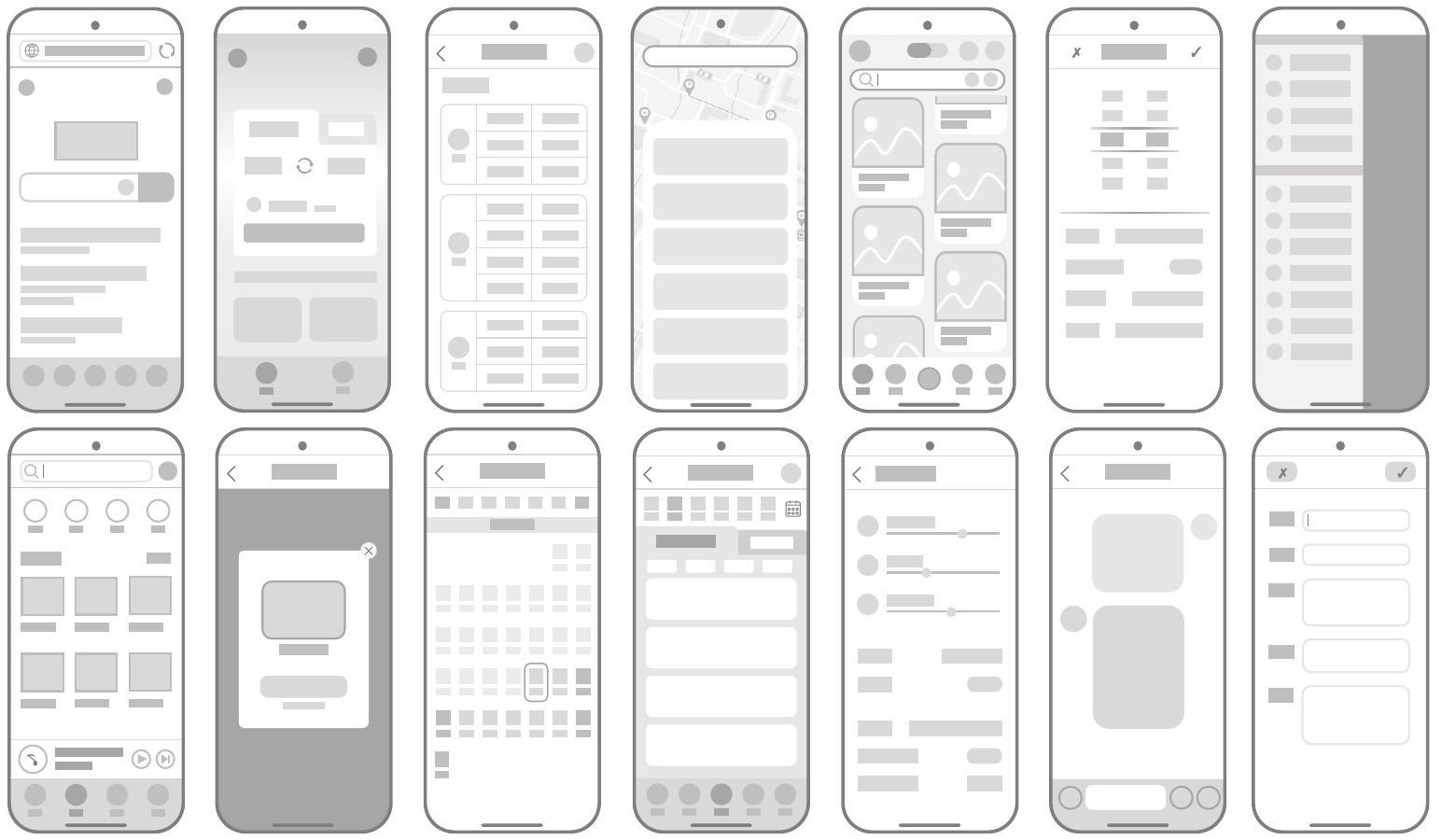}
    \caption{Typical UI layouts of 136 screenshots.}
    \label{fig:layout}
\end{figure}

Figure~\ref{fig:layout} displays typical UI layouts for the 136 screenshots used in the first experiment, as described in Sec.~\ref{sec:7_1}.

\section{Questions Examples}
Table~\ref{tab:question_exp} presents sample questions for the five categories of UI question answering tasks used in first experiment of Sec.~\ref{sec:7_1}.

\begin{table}[htbp]
    \centering
    \scriptsize
    \renewcommand{\arraystretch}{1.0}
    \caption{Example Questions used in Sec.~\ref{sec:7_1}.}
    \begin{tabular}{lll}
        \toprule
        Category & \# & Example questions\\
        \midrule
        Status inquiry & 89 & Which file has been selected?\\
        && Is flight mode turned on?\\
        && Which item in the bottom navigation bar is selected?\\
        && Which date has been selected?\\
        && Is the vibration feature enabled?\\
        \midrule
        Content question & 188 & What is the customer service phone number?\\
        && Where is the current location?\\
        && How many parcels have been sent out?\\
        && What is the title of the second news article?\\
        && What is the rating of this restaurant?\\
        \midrule
        Item count&68& How many flights are displayed? What are they?\\
        && How many travel plans are available? What are they?\\
        && How many types of filters does the UI offer? What are they?\\
        && How many to-do items are there? What are they?\\
        && How many fields need to be filled out? What are they?\\
        \midrule
        Functionality query&105&How can I become a member?\\
        && How can I change the set work hours?\\
        && How can I increase the number of rooms booked?\\
        && How can I delete answers that I'm not interested in?\\
        && How can I check the balance of my mobile phone charges?\\
        \midrule
        Purpose summary&136& What are the main functions or purposes of the current UI?\\
        \bottomrule
    \end{tabular}    
    \label{tab:question_exp}
\end{table}

\section{List of 147 Real Tasks}
\label{sec:tasks}
\textcolor{black}{Table~\ref{tab:147} presents the 147 actual tasks utilized in the third experiment as detailed in Section~\ref{sec:7_3}. The table includes details such as the originating app for each task, its category, description, and the number of human evaluators able to complete each task.
}



\aptLtoX[graphic=no,type=html]{\begin{table}
\tiny
\captionof{table}{List of 147 real tasks in Section~\ref{sec:7_3}, including the source app, the task category, description, and the number of human evaluators who were able to complete the task.} 
\label{tab:147}
\begin{tabular}{|>{\centering}p{1.0cm}|>{\centering}p{1.0cm}|p{5.5cm}|p{0.3cm}|}
   \hline
    \textbf{Category} & \textbf{App} & \textbf{Task} & \textbf{\#} \\ 
    \hline
        Media & Bilibili & Play ``Guardians of Liberation West 4'' on Bilibili. & 12\\ 
        ~ & ~ & Check offline-cached videos on Bilibili. & 12 \\ 
        ~ & ~ & Cache the first episode of ``Guardians of Liberation West 4''. & 12 \\ 
        ~ & ~ & Buy an invitation code on Bilibili. & 2 \\ 
        ~ & ~ & Check my manuscript ``Test'' in Bilibili's creation center. & 12 \\ 
        ~ & QQ Music & Play ``Ten Years'' by Eason Chan on QQ Music and favorite it. & 12 \\ 
        ~ & ~ & Add the acoustic version of ``Don't Cry, Girl'' to the self-created playlist ``Test'' on QQ Music. & 12 \\ 
        ~ & ~ & Set a 45-minute timer for auto shutdown on QQ Music. & 12 \\ 
        ~ & ~ & Check my favorite songs on QQ Music. & 12 \\ 
        ~ & Spotify & Search for popular Chinese music playlists on Spotify.  & 12\\ 
        ~ & ~ & View my playlists on Spotify. & 12 \\ 
        ~ & ~ & Play songs by the artist group ``Red Velvet'' on Spotify. & 12\\ 
        ~ & ~ & View and play my liked songs on Spotify.  & 12\\ 
        ~ & Ximalaya & Play a children's bedtime story on Ximalaya. & 12 \\ 
        ~ & ~ & Set a timer to stop playing after 3 episodes on Ximalaya.  & 9\\ 
        ~ & ~ & Check my expired coupons on Ximalaya. & 9 \\ 
        ~ & ~ & View content I've purchased on Ximalaya.  & 12\\ 
        ~ & Bili. Comics & Claim the comic welfare voucher for Bilibili Comics' big member.  & 1\\ 
        ~ & ~ & Disable the automatic purchase of the next chapter on Bilibili Comics. & 12 \\ 
        ~ & ~ & Watch the cached comics on Bilibili Comics. & 9 \\ 
        ~ & ~ & View my purchased comics on Bilibili Comics.  & 12\\ 
        ~ & ~ & Check the Chinese comic ranking on Bilibili Comics.  & 12\\ 
        \hline
        Social & WeChat & Reminding a friend to have breakfast with a Wechat message. & 12 \\ 
        ~ & ~ & Check my WeChat Moments.  & 12\\ 
        ~ & ~ & Post the newly taken picture on WeChat Moments texting ``Morning''. & 12\\ 
        ~ & ~ & Enter a chat on WeChat and add the ``Line Puppy Vol.1'' emoji. & 10 \\ 
        ~ & Xiaohongshu & Search for men's winter outfits on Xiaohongshu and like the first post.  & 12\\ 
        ~ & ~ & Check my browsing history on Xiaohongshu. & 11 \\ 
        ~ & ~ & View coupons on Xiaohongshu.  & 9\\ 
        ~ & ~ & View gift-receiving records on Xiaohongshu.  & 3\\ 
        ~ & Weibo & Find and follow the official Weibo account of a local institute.  & 12\\ 
        ~ & ~ & View all messages that mention me on Weibo. & 12 \\ 
        ~ & ~ & Check Weibo's trending topics.  & 12\\ 
        ~ & ~ & View comments I've made on Weibo.  & 10\\ 
        \hline
        Travel & Amap & Use Amap to find the route to a local station by bus and subway.  & 12\\ 
        ~ & ~ & View satellite maps on Amap.  & 10\\ 
        ~ & ~ & Search for and bookmark a local station on Amap.  & 12\\ 
        ~ & ~ & Note the favorite station as a high-speed railway station on Amap.  & 11\\ 
        ~ & Ctrip & Book a guesthouse near a local spot from Dec. 14th to 15th on Ctrip.  & 12\\ 
        ~ & ~ & Check my upcoming travel orders on Ctrip.  & 12\\ 
        ~ & ~ & Explore tourist attractions in a certain city on Ctrip.  & 12\\ 
        ~ & ~ & Import flight tickets booked by others on Ctrip.  & 1\\ 
        ~ & ~ & View my travel itinerary on Ctrip.  & 12\\ 
        ~ &  12306 & View the station display on 12306.  & 12\\ 
        ~ & ~ & Apply for a temporary identity certificate on 12306. & 12 \\ 
        ~ & ~ & Search and call the customer service of a city on 12306.  & 12\\ 
        ~ & ~ & View pending evaluations for catering orders on 12306.  & 11\\ 
        ~ & ~ & Go to 12306 to check tickets from a city to another city on Dec. 27th.  & 12\\ 
        ~ & Umetrip & Check the airport display at Shanghai Pudong Airport on Umetrip.  & 10\\ 
        ~ & ~ & Check my upcoming travel itinerary on Umetrip.  & 12\\ 
        ~ & ~ & View my benefits package on Umetrip. & 12 \\ 
        ~ & ~ & Add personal passport information on Umetrip.  & 10\\ 
        \hline
        Shopping & Taobao & View my favorite items on Taobao.  & 12\\ 
        ~ & ~ & Delete the first refund record on Taobao orders.  & 10\\ 
        ~ & ~ & Search for Logitech keyboards on Taobao and select the first item. & 12\\ 
        ~ & ~ & Add price protection to all orders on Taobao.  & 1\\ 
        ~ & Freshippo & Empty the shopping cart on Freshippo.  & 12\\ 
        ~ & ~ & Select all items in the Freshippo shopping cart and delete them. & 12 \\ 
        ~ & ~ & Buy hairy crabs on Freshippo.  & 12\\ 
        ~ & ~ & View my invoice history on Freshippo.  & 12\\ 
        ~ & ~ & Add a shipping address on Freshippo. & 12 \\ 
        ~ & Pinduoduo & Search for hair accessories on Pinduoduo and select a branded item.  & 12\\ 
        ~ & ~ & Check pending refund orders on Pinduoduo.  & 12\\ 
        ~ & ~ & Check the progress of my reported complaints on Pinduoduo.  & 3\\ 
        ~ & ~ & View my favorite items on Pinduoduo.  & 12\\ 
        \hline
        Photography & Meitu & Take a photo on Meitu Xiuxiu and add the filter ``Midnight Diner.''  & 12\\ 
        ~ & ~ & Create a one-inch photo on Meitu Xiuxiu. & 4 \\ 
        ~ & ~ & Set my personal watermark on Meitu Xiuxiu.  & 5\\ 
        ~ & ~ & Disallow others from saving my photos on Meitu Xiuxiu. & 5 \\ 
        ~ & Qingyan & View the effects I've used on Qingyan.  & 8\\ 
        ~ & ~ & Turn on the flashlight in Qingyan.  & 12\\                 
        ~ & ~ & Turn off the male makeup adaptation on Qingyan.  & 9\\ 
        \hline
        Finance & Alipay & Withdraw balance from Alipay.  & 12\\ 
        ~ & ~ & Check my Alipay monthly bill expenses.  & 12\\ 
        ~ & ~ & Cancel password-free payments on Alipay for Taobao.  & 10\\ 
        ~ & WeChat & View the balance in WeChat.  & 12\\ 
        ~ & ~ & Pay heating fees using WeChat.  & 4\\ 
        \hline
        Food & Ele.me & Add a new delivery address on Ele.me.  & 12\\ 
        ~ & ~ & Check fried chicken delivery options on Ele.me, sorted by sales volume.  & 12\\ 
        ~ & ~ & View my available coupons on Ele.me.  & 12\\ 
        ~ & ~ & Enable small-amount contactless payment on Ele.me.  & 5\\ 
        ~ & Meituan & Place an order for braised chicken rice at a restaurant on Meituan.  & 12\\ 
        ~ & ~ & Check my refunded orders on Meituan.  & 12\\ 
        ~ & ~ & View my favorites on Meituan.  & 12\\ 
        ~ & ~ & Explore KTV options within a 3 km radius on Meituan.  & 12\\ 
        ~ & Damai & Add a new attendee for events on Damai. & 11\\ 
        ~ & ~ & Save the official e-ticket for a DeYunShe performance I attended on Damai to my photo album.  & 1\\ 
        ~ & ~ & View my past event tickets on Damai.  & 12\\ 
        ~ & ~ & Check nearby stand-up comedy show tickets on Damai.  & 12\\ 
        ~ & ~ & Enable point expiration reminders on Damai.  & 3\\ 
        \hline
        Health & Keep & Check courses in the ``Ba Duan Jin'' series on Keep.  & 12\\ 
        ~ & ~ & Start recording walking activities on Keep.  & 10\\ 
        ~ & ~ & View information about my sports equipment on Keep.  & 12\\ 
        ~ & ~ & Enable posture assessment on Keep.  & 11\\  
\hline
 \multicolumn{4}{|r|}{\tiny\sl continued on next page} \\
\bottomrule
\end{tabular}
\end{table}

 \begin{table}
\tiny
\begin{tabular}{|>{\centering}p{1.2cm}|>{\centering}p{1.0cm}|p{5.1cm}|p{0.3cm}|}
\hline
\multicolumn{4}{|l|}{\tiny\sl continued from previous page} \\
   \hline
    \textbf{Category} & \textbf{App} & \textbf{Task} & \textbf{\#} \\ 
    \hline
        ~ & Xunji & View exercises for biceps in Xunji.  & 12\\ 
        ~ & ~ & Check my body data in Xunji.  & 12\\ 
        ~ & ~ & Play the tutorial for the ``Bird Fly'' exercise in Xunji. & 12\\ 
        ~ & ~ & View training statistics in Xunji. & 11\\ 
        \hline
        Learning & Youdao & Translate ``deep learning'' using Youdao. & 12\\ 
        ~ & ~ & Memorize today's vocabulary using Youdao. & 11\\ 
        ~ & ~ & Download the offline Oxford dictionary from Youdao. & 4\\ 
        ~ & Toutiao & Check the top headlines on Toutiao. & 12\\ 
        ~ & ~ & View my browsing history on Toutiao. & 12\\ 
        ~ & ~ & Open news notifications on Toutiao. & 11\\ 
        ~ & Weread & Add ``Journey to the West'' to my bookshelf on Weread. & 12\\ 
        ~ & ~ & Check my reading duration on Weread. & 12\\ 
        ~ & ~ & Set permission to use volume keys for page turning on Weread. & 9\\ 
        ~ & Zhihu & Search for ``CS ranking in China'' on Zhihu and unfold the first answer. & 12\\ 
        ~ & ~ & View my saved content on Zhihu. & 12\\ 
        ~ & ~ & View content I've liked on Zhihu. & 4\\ 
        \hline
        Productivity & NetEase Mail & Mark all unread emails as read in NetEase Mail Master. & 11\\ 
        ~ & Ten. Meeting & Initiate a quick meeting on Tencent Meeting, using video and PMI. & 12\\ 
        ~ & Calendar & Add a reminder for ``Go for class'' at 2:00 PM today on Calendar. & 12\\ 
        ~ & ~ & Add a reminder for ``Buy gifts'' on Christmas on Calendar. & 12\\ 
        ~ & QQ Mail & Open the inbox on QQ Mail and view the daily reading content. & 11\\ 
        ~ & ~ & Write an email to a certain email on QQ Mail. & 12\\ 
        ~ & Messaging & View notification messages in the Messages app. & 11\\ 
        ~ & ~ & Delete all spam messages in the messages app. & 8\\ 
        ~ & Phone & Call Alibaba DingTalk customer service from contacts. & 12\\ 
        ~ & Contacts & Share the QR code of a friend in Contacts with WeChat file assistant. & 10\\ 
        ~ & ~ & Set the workplace of a certain friend to a local institute in Contacts. & 12\\ 
        ~ & Notepad & Add a ``Go to the dentist on Sunday'' task in the Notepad. & 11\\ 
        ~ & ~ & View my favorite notes in the Notepad. & 10\\ 
        ~ & Chrome & Open an incognito window in Chrome. & 12\\ 
        ~ & ~ & Clear cookies data in Chrome. & 11\\ 
        ~ & ~ & Set Bing as the default search engine in Chrome. & 12\\ 
        \hline
        Living & China Mobile & Top up mobile credit with China Mobile. & 12\\ 
        ~ & ~ & Check my subscribed plans with China Mobile. & 12\\ 
        ~ & Cainiao & Check how many packages are waiting for pickup with Cainiao. & 12\\ 
        ~ & ~ & Add a family member account in Cainiao. & 8\\ 
        ~ & ~ & Enable package pickup authorization for Fengchao in the Cainiao. & 3\\ 
        ~ & Alipay & Help me pay a 10 CNY electricity bill in Alipay. & 11\\ 
        ~ & ~ & Open the public transportation code in Alipay. & 12\\ 
        ~ & ~ & Recharge my campus card in Alipay's ``Campus'' mini-program. & 12\\ 
        ~ & Weather & Check the weather in Beijing in the system's Weather app. & 12\\ 
        ~ & Clock & Set a new alarm at 9:30 AM in the system's Clock app. & 12\\ 
        ~ & ~ & Check the current time in New York in the system's Clock app. & 12\\ 
        ~ & Local App & Book a gym session at the app of a local institute. & 7\\ 
        ~ & ~ & Check my safety mailbox at the app of a local institute. & 11\\ 
        \hline
        Settings & Settings & Go to Settings, open Bluetooth, and connect to AirPods. & 12\\ 
        ~ & ~ & Go to Settings and enable the personal mobile WLAN hotspot. & 12\\ 
        ~ & ~ & Go to Settings and set the time format to 24-hour. & 11\\ 
        ~ & Alipay & Set dark mode in Alipay to follow the system. & 11\\ 
        ~ & ~ & Check the email information associated with the Alipay account. & 11\\ 
        ~ & ~ & Go to Alipay and disable adding a friend through the transfer page. & 5\\ 
        ~ & WeChat & Change the WeChat interface language to English. & 12\\ 
        ~ & ~ & Go to WeChat and add the ``Search'' feature to the Discover page.  & 9\\
\bottomrule
\end{tabular}

 \end{table}}{\clearpage
\tiny
\tablefirsthead{
    \hline
    \textbf{Category} & \textbf{App} & \textbf{Task} & \textbf{\#} \\ 
    \hline
    }
\tablehead{
    \hline 
    \multicolumn{4}{|l|}{\tiny\sl continued from previous page} \\ 
    \hline
    \textbf{Category} & \textbf{App} & \textbf{Task} & \textbf{\#} \\ 
    \hline
    }
\tabletail{
    \hline
    \multicolumn{4}{|r|}{\tiny\sl continued on next page} \\
    \hline
    }
\tablelasttail{\hline}
\captionof{table}{List of 147 real tasks in Section~\ref{sec:7_3}, including the source app, the task category, description, and the number of human evaluators who were able to complete the task.} 
\label{tab:147}
\begin{xtabular}{|>{\centering}p{0.9cm}|>{\centering}p{1.0cm}|p{4.9cm}|p{0.2cm}|}
        Media & Bilibili & Play ``Guardians of Liberation West 4'' on Bilibili. & 12\\ 
        ~ & ~ & Check offline-cached videos on Bilibili. & 12 \\ 
        ~ & ~ & Cache the first episode of ``Guardians of Liberation West 4''. & 12 \\ 
        ~ & ~ & Buy an invitation code on Bilibili. & 2 \\ 
        ~ & ~ & Check my manuscript ``Test'' in Bilibili's creation center. & 12 \\ 
        ~ & QQ Music & Play ``Ten Years'' by Eason Chan on QQ Music and favorite it. & 12 \\ 
        ~ & ~ & Add the acoustic version of ``Don't Cry, Girl'' to the self-created playlist ``Test'' on QQ Music. & 12 \\ 
        ~ & ~ & Set a 45-minute timer for auto shutdown on QQ Music. & 12 \\ 
        ~ & ~ & Check my favorite songs on QQ Music. & 12 \\ 
        ~ & Spotify & Search for popular Chinese music playlists on Spotify.  & 12\\ 
        ~ & ~ & View my playlists on Spotify. & 12 \\ 
        ~ & ~ & Play songs by the artist group ``Red Velvet'' on Spotify. & 12\\ 
        ~ & ~ & View and play my liked songs on Spotify.  & 12\\ 
        ~ & Ximalaya & Play a children's bedtime story on Ximalaya. & 12 \\ 
        ~ & ~ & Set a timer to stop playing after 3 episodes on Ximalaya.  & 9\\ 
        ~ & ~ & Check my expired coupons on Ximalaya. & 9 \\ 
        ~ & ~ & View content I've purchased on Ximalaya.  & 12\\ 
        ~ & Bili. Comics & Claim the comic welfare voucher for Bilibili Comics' big member.  & 1\\ 
        ~ & ~ & Disable the automatic purchase of the next chapter on Bilibili Comics. & 12 \\ 
        ~ & ~ & Watch the cached comics on Bilibili Comics. & 9 \\ 
        ~ & ~ & View my purchased comics on Bilibili Comics.  & 12\\ 
        ~ & ~ & Check the Chinese comic ranking on Bilibili Comics.  & 12\\ 
        \hline
        Social & WeChat & Reminding a friend to have breakfast with a Wechat message. & 12 \\ 
        ~ & ~ & Check my WeChat Moments.  & 12\\ 
        ~ & ~ & Post the newly taken picture on WeChat Moments texting ``Morning''. & 12\\ 
        ~ & ~ & Enter a chat on WeChat and add the ``Line Puppy Vol.1'' emoji. & 10 \\ 
        ~ & Xiaohongshu & Search for men's winter outfits on Xiaohongshu and like the first post.  & 12\\ 
        ~ & ~ & Check my browsing history on Xiaohongshu. & 11 \\ 
        ~ & ~ & View coupons on Xiaohongshu.  & 9\\ 
        ~ & ~ & View gift-receiving records on Xiaohongshu.  & 3\\ 
        ~ & Weibo & Find and follow the official Weibo account of a local institute.  & 12\\ 
        ~ & ~ & View all messages that mention me on Weibo. & 12 \\ 
        ~ & ~ & Check Weibo's trending topics.  & 12\\ 
        ~ & ~ & View comments I've made on Weibo.  & 10\\ 
        \hline
        Travel & Amap & Use Amap to find the route to a local station by bus and subway.  & 12\\ 
        ~ & ~ & View satellite maps on Amap.  & 10\\ 
        ~ & ~ & Search for and bookmark a local station on Amap.  & 12\\ 
        ~ & ~ & Note the favorite station as a high-speed railway station on Amap.  & 11\\ 
        ~ & Ctrip & Book a guesthouse near a local spot from Dec. 14th to 15th on Ctrip.  & 12\\ 
        ~ & ~ & Check my upcoming travel orders on Ctrip.  & 12\\ 
        ~ & ~ & Explore tourist attractions in a certain city on Ctrip.  & 12\\ 
        ~ & ~ & Import flight tickets booked by others on Ctrip.  & 1\\ 
        ~ & ~ & View my travel itinerary on Ctrip.  & 12\\ 
        ~ &  12306 & View the station display on 12306.  & 12\\ 
        ~ & ~ & Apply for a temporary identity certificate on 12306. & 12 \\ 
        ~ & ~ & Search and call the customer service of a city on 12306.  & 12\\ 
        ~ & ~ & View pending evaluations for catering orders on 12306.  & 11\\ 
        ~ & ~ & Go to 12306 to check tickets from a city to another city on Dec. 27th.  & 12\\ 
        ~ & Umetrip & Check the airport display at Shanghai Pudong Airport on Umetrip.  & 10\\ 
        ~ & ~ & Check my upcoming travel itinerary on Umetrip.  & 12\\ 
        ~ & ~ & View my benefits package on Umetrip. & 12 \\ 
        ~ & ~ & Add personal passport information on Umetrip.  & 10\\ 
        \hline
        Shopping & Taobao & View my favorite items on Taobao.  & 12\\ 
        ~ & ~ & Delete the first refund record on Taobao orders.  & 10\\ 
        ~ & ~ & Search for Logitech keyboards on Taobao and select the first item. & 12\\ 
        ~ & ~ & Add price protection to all orders on Taobao.  & 1\\ 
        ~ & Freshippo & Empty the shopping cart on Freshippo.  & 12\\ 
        ~ & ~ & Select all items in the Freshippo shopping cart and delete them. & 12 \\ 
        ~ & ~ & Buy hairy crabs on Freshippo.  & 12\\ 
        ~ & ~ & View my invoice history on Freshippo.  & 12\\ 
        ~ & ~ & Add a shipping address on Freshippo. & 12 \\ 
        ~ & Pinduoduo & Search for hair accessories on Pinduoduo and select a branded item.  & 12\\ 
        ~ & ~ & Check pending refund orders on Pinduoduo.  & 12\\ 
        ~ & ~ & Check the progress of my reported complaints on Pinduoduo.  & 3\\ 
        ~ & ~ & View my favorite items on Pinduoduo.  & 12\\ 
        \hline
        Photography & Meitu & Take a photo on Meitu Xiuxiu and add the filter ``Midnight Diner.''  & 12\\ 
        ~ & ~ & Create a one-inch photo on Meitu Xiuxiu. & 4 \\ 
        ~ & ~ & Set my personal watermark on Meitu Xiuxiu.  & 5\\ 
        ~ & ~ & Disallow others from saving my photos on Meitu Xiuxiu. & 5 \\ 
        ~ & Qingyan & View the effects I've used on Qingyan.  & 8\\ 
        ~ & ~ & Turn on the flashlight in Qingyan.  & 12\\                 
        ~ & ~ & Turn off the male makeup adaptation on Qingyan.  & 9\\ 
        \hline
        Finance & Alipay & Withdraw balance from Alipay.  & 12\\ 
        ~ & ~ & Check my Alipay monthly bill expenses.  & 12\\ 
        ~ & ~ & Cancel password-free payments on Alipay for Taobao.  & 10\\ 
        ~ & WeChat & View the balance in WeChat.  & 12\\ 
        ~ & ~ & Pay heating fees using WeChat.  & 4\\ 
        \hline
        Food & Ele.me & Add a new delivery address on Ele.me.  & 12\\ 
        ~ & ~ & Check fried chicken delivery options on Ele.me, sorted by sales volume.  & 12\\ 
        ~ & ~ & View my available coupons on Ele.me.  & 12\\ 
        ~ & ~ & Enable small-amount contactless payment on Ele.me.  & 5\\ 
        ~ & Meituan & Place an order for braised chicken rice at a restaurant on Meituan.  & 12\\ 
        ~ & ~ & Check my refunded orders on Meituan.  & 12\\ 
        ~ & ~ & View my favorites on Meituan.  & 12\\ 
        ~ & ~ & Explore KTV options within a 3 km radius on Meituan.  & 12\\ 
        ~ & Damai & Add a new attendee for events on Damai. & 11\\ 
        ~ & ~ & Save the official e-ticket for a DeYunShe performance I attended on Damai to my photo album.  & 1\\ 
        ~ & ~ & View my past event tickets on Damai.  & 12\\ 
        ~ & ~ & Check nearby stand-up comedy show tickets on Damai.  & 12\\ 
        ~ & ~ & Enable point expiration reminders on Damai.  & 3\\ 
        \hline
        Health & Keep & Check courses in the ``Ba Duan Jin'' series on Keep.  & 12\\ 
        ~ & ~ & Start recording walking activities on Keep.  & 10\\ 
        ~ & ~ & View information about my sports equipment on Keep.  & 12\\ 
        ~ & ~ & Enable posture assessment on Keep.  & 11\\  
        ~ & Xunji & View exercises for biceps in Xunji.  & 12\\ 
        ~ & ~ & Check my body data in Xunji.  & 12\\ 
        ~ & ~ & Play the tutorial for the ``Bird Fly'' exercise in Xunji. & 12\\ 
        ~ & ~ & View training statistics in Xunji. & 11\\ 
        \hline
        Learning & Youdao & Translate ``deep learning'' using Youdao. & 12\\ 
        ~ & ~ & Memorize today's vocabulary using Youdao. & 11\\ 
        ~ & ~ & Download the offline Oxford dictionary from Youdao. & 4\\ 
        ~ & Toutiao & Check the top headlines on Toutiao. & 12\\ 
        ~ & ~ & View my browsing history on Toutiao. & 12\\ 
        ~ & ~ & Open news notifications on Toutiao. & 11\\ 
        ~ & Weread & Add ``Journey to the West'' to my bookshelf on Weread. & 12\\ 
        ~ & ~ & Check my reading duration on Weread. & 12\\ 
        ~ & ~ & Set permission to use volume keys for page turning on Weread. & 9\\ 
        ~ & Zhihu & Search for ``CS ranking in China'' on Zhihu and unfold the first answer. & 12\\ 
        ~ & ~ & View my saved content on Zhihu. & 12\\ 
        ~ & ~ & View content I've liked on Zhihu. & 4\\ 
        \hline
        Productivity & NetEase Mail & Mark all unread emails as read in NetEase Mail Master. & 11\\ 
        ~ & Ten. Meeting & Initiate a quick meeting on Tencent Meeting, using video and PMI. & 12\\ 
        ~ & Calendar & Add a reminder for ``Go for class'' at 2:00 PM today on Calendar. & 12\\ 
        ~ & ~ & Add a reminder for ``Buy gifts'' on Christmas on Calendar. & 12\\ 
        ~ & QQ Mail & Open the inbox on QQ Mail and view the daily reading content. & 11\\ 
        ~ & ~ & Write an email to a certain email on QQ Mail. & 12\\ 
        ~ & Messaging & View notification messages in the Messages app. & 11\\ 
        ~ & ~ & Delete all spam messages in the messages app. & 8\\ 
        ~ & Phone & Call Alibaba DingTalk customer service from contacts. & 12\\ 
        ~ & Contacts & Share the QR code of a friend in Contacts with WeChat file assistant. & 10\\ 
        ~ & ~ & Set the workplace of a certain friend to a local institute in Contacts. & 12\\ 
        ~ & Notepad & Add a ``Go to the dentist on Sunday'' task in the Notepad. & 11\\ 
        ~ & ~ & View my favorite notes in the Notepad. & 10\\ 
        ~ & Chrome & Open an incognito window in Chrome. & 12\\ 
        ~ & ~ & Clear cookies data in Chrome. & 11\\ 
        ~ & ~ & Set Bing as the default search engine in Chrome. & 12\\ 
        \hline
        Living & China Mobile & Top up mobile credit with China Mobile. & 12\\ 
        ~ & ~ & Check my subscribed plans with China Mobile. & 12\\ 
        ~ & Cainiao & Check how many packages are waiting for pickup with Cainiao. & 12\\ 
        ~ & ~ & Add a family member account in Cainiao. & 8\\ 
        ~ & ~ & Enable package pickup authorization for Fengchao in the Cainiao. & 3\\ 
        ~ & Alipay & Help me pay a 10 CNY electricity bill in Alipay. & 11\\ 
        ~ & ~ & Open the public transportation code in Alipay. & 12\\ 
        ~ & ~ & Recharge my campus card in Alipay's ``Campus'' mini-program. & 12\\ 
        ~ & Weather & Check the weather in Beijing in the system's Weather app. & 12\\ 
        ~ & Clock & Set a new alarm at 9:30 AM in the system's Clock app. & 12\\ 
        ~ & ~ & Check the current time in New York in the system's Clock app. & 12\\ 
        ~ & Local App & Book a gym session at the app of a local institute. & 7\\ 
        ~ & ~ & Check my safety mailbox at the app of a local institute. & 11\\ 
        \hline
        Settings & Settings & Go to Settings, open Bluetooth, and connect to AirPods. & 12\\ 
        ~ & ~ & Go to Settings and enable the personal mobile WLAN hotspot. & 12\\ 
        ~ & ~ & Go to Settings and set the time format to 24-hour. & 11\\ 
        ~ & Alipay & Set dark mode in Alipay to follow the system. & 11\\ 
        ~ & ~ & Check the email information associated with the Alipay account. & 11\\ 
        ~ & ~ & Go to Alipay and disable adding a friend through the transfer page. & 5\\ 
        ~ & WeChat & Change the WeChat interface language to English. & 12\\ 
        ~ & ~ & Go to WeChat and add the ``Search'' feature to the Discover page.  & 9\\
\end{xtabular}}

\end{document}